\documentclass[a4paper]{article}
\usepackage{eurosym}
\usepackage{amssymb}
\usepackage{amsfonts}
\usepackage{amsmath}
\usepackage[margin=2cm]{geometry}
\usepackage{graphicx}
\usepackage{caption}
\usepackage{subfig}
\usepackage{multirow,makecell}
\usepackage{multicol}
\usepackage{hhline}
\usepackage{stackengine}
\usepackage{array}
\usepackage{url}

\newcommand\xrowht[2][0]{\addstackgap[.5\dimexpr#2\relax]{\vphantom{#1}}}

\input{tcilatex}

\begin{document}

\title{Energy dissipation for hereditary and \\
energy conservation for non-local fractional wave equations}
\author{Du\v{s}an Zorica\thanks{
Mathematical Institute, Serbian Academy of Arts and Sciences, Kneza Mihaila
36, 11000 Belgrade, Serbia and Department of Physics, Faculty of Sciences,
University of Novi Sad, Trg D. Obradovi\'{c}a 4, 21000 Novi Sad, Serbia,
dusan\textunderscore zorica@mi.sanu.ac.rs}, Ljubica Oparnica\thanks{
Faculty of Education, University of Novi Sad, Podgori\v cka 4, 25000 Sombor,
Serbia and Department of Mathematics: Analysis, Logic and Discrete
Mathematics, University of Gent, Krijgslaan 281 (building S8), 9000 Gent,
Belgium, Oparnica.Ljubica@UGent.be}}
\maketitle

\begin{abstract}
\noindent Using the method of a priori energy estimates, energy dissipation
is proved for the class of hereditary fractional wave equations, obtained
through the system of equations consisting of equation of motion, strain,
and fractional order constitutive models, that include the distributed-order
constitutive law in which the integration is performed from zero to one
generalizing all linear constitutive models of fractional and integer
orders, as well as for the thermodynamically consistent fractional Burgers
models, where the orders of fractional differentiation are up to the second
order. In the case of non-local fractional wave equations, obtained using
non-local constitutive models of Hooke- and Eringen-type in addition to the
equation of motion and strain, a priori energy estimates yield the energy
conservation, with the reinterpreted notion of the potential energy.

\end{abstract}

\section{Introduction}

Fractional wave equations, describing the disturbance propagation in a
viscoelastic or non-local material, are obtained through the system of
equations consisting of: equation of motion corresponding to one-dimensional
deformable body%
\begin{equation}
\partial _{x}\sigma (x,t)=\rho \,\partial _{tt}u(x,t),  \label{eq-motion}
\end{equation}%
where $u$ and $\sigma $ are displacement and stress, assumed as functions of
space $x\in 
\mathbb{R}
$ and time $t>0,$ with $\rho $ being constant material density, strain for
small local deformations%
\begin{equation}
\varepsilon (x,t)=\partial _{x}u(x,t),  \label{strejn}
\end{equation}%
and constitutive equation connecting stress and strain, which can model
either hereditary or non-local material properties.

The aim is to investigate the energy conserving properties of such obtained
wave equations, namely hereditary and non-local wave equations. Hereditary
materials are modelled by the fractional-order constitutive equations of
viscoelastic body including distributed-order model containing fractional
differentiation orders up to the first order, as well as fractional Burgers
models containing also the differentiation orders up to the second order.
Energy dissipation is expected for hereditary wave equations, since the
thermodynamical requirements on the model parameters impose dissipativity of
such constitutive models. On the other hand, non-local materials, modelled
by the non-local Hooke law and fractional Eringen stress gradient model, are
not expected to dissipate energy.

Hereditary effects in a viscoelastic body are modelled either by the
distributed-order constitutive equation%
\begin{equation}
\int_{0}^{1}\phi _{\sigma }(\alpha )\,{}_{0}\mathrm{D}_{t}^{\alpha }\sigma
(x,t)\,\mathrm{d}\alpha =\int_{0}^{1}\phi _{\varepsilon }(\alpha )\,{}_{0}%
\mathrm{D}_{t}^{\alpha }\varepsilon (x,t)\,\mathrm{d}\alpha ,
\label{const-eq}
\end{equation}%
where $\phi _{\sigma }$ and $\phi _{\varepsilon }$ are constitutive
functions or distributions and where fractional differentiation orders do
not exceed the first order, or by the thermodynamically consistent
fractional Burgers models where fractional differentiation orders are up to
the second order. The fractional Burgers models are represented by unified
models belonging to two classes: the first class is represented by the
unified constitutive equation%
\begin{equation}
\left( 1+a_{1}\,{}_{0}\mathrm{D}_{t}^{\alpha }+a_{2}\,{}_{0}\mathrm{D}%
_{t}^{\beta }+a_{3}\,{}_{0}\mathrm{D}_{t}^{\gamma }\right) \sigma \left(
x,t\right) =\left( b_{1}\,{}_{0}\mathrm{D}_{t}^{\mu }+b_{2}\,{}_{0}\mathrm{D}%
_{t}^{\mu +\eta }\right) \varepsilon \left( x,t\right) ,  \label{UCE-1-5}
\end{equation}%
while the second one is represented by%
\begin{equation}
\left( 1+a_{1}\,{}_{0}\mathrm{D}_{t}^{\alpha }+a_{2}\,{}_{0}\mathrm{D}%
_{t}^{\beta }+a_{3}\,{}_{0}\mathrm{D}_{t}^{\beta +\eta }\right) \sigma
\left( x,t\right) =\left( b_{1}\,{}_{0}\mathrm{D}_{t}^{\beta }+b_{2}\,{}_{0}%
\mathrm{D}_{t}^{\beta +\eta }\right) \varepsilon \left( x,t\right) ,
\label{UCE-6-8}
\end{equation}%
where $a_{1},a_{2},a_{3},b_{1},b_{2}>0,$ $\alpha ,\beta ,\mu \in \left[ 0,1%
\right] ,$ with $\alpha \leq \beta ,$ $\gamma \in \left[ 0,2\right] ,$ and $%
\eta \in \left\{ \alpha ,\beta \right\} .$ The operator of Riemann-Liouville
fractional derivative ${}_{0}\mathrm{D}_{t}^{\xi }$ of order $\xi \in \left[
n,n+1\right] ,$ $n\in 
\mathbb{N}
_{0},$ used in constitutive models (\ref{const-eq}), (\ref{UCE-1-5}), and (%
\ref{UCE-6-8}), is defined by%
\begin{equation*}
{}_{0}\mathrm{D}_{t}^{\xi }y\left( t\right) =\frac{\mathrm{d}^{n+1}}{\mathrm{%
d}t^{n+1}}\left( \frac{t^{-\left( \xi -n\right) }}{\Gamma \left( 1-\left(
\xi -n\right) \right) }\ast _{t}y\left( t\right) \right) ,\;\;t>0,
\end{equation*}%
see \cite{TAFDE}, where $\ast _{t}$ denotes the convolution in time: $%
f\left( t\right) \ast _{t}g\left( t\right) =\int_{0}^{t}f\left( t^{\prime
}\right) g\left( t-t^{\prime }\right) \mathrm{d}t^{\prime },$ $t>0.$

Non-locality effects in a material are described either by the non-local
Hooke law%
\begin{equation}
\sigma (x,t)=\frac{E}{\ell ^{1-\alpha }}\frac{|x|^{-\alpha }}{2\Gamma
(1-\alpha )}\ast _{x}\varepsilon (x,t),\;\;\alpha \in \left( 0,1\right) ,
\label{nl-Huk}
\end{equation}%
or by the fractional Eringen constitutive equation 
\begin{equation}
\sigma \left( x,t\right) -\ell ^{\alpha }\,\mathrm{D}_{x}^{\alpha }\sigma
\left( x,t\right) =E\,\varepsilon \left( x,t\right) ,\;\;\alpha \in \left(
1,3\right) ,  \label{frac-Eringen}
\end{equation}%
where $E$ is Young modulus, $\ell $ is non-locality parameter, and $\mathrm{D%
}_{x}^{\alpha }$ is defined as%
\begin{eqnarray}
\mathrm{D}_{x}^{\alpha }y\left( x\right) &=&\frac{\left\vert x\right\vert
^{1-\alpha }}{2\Gamma \left( 2-\alpha \right) }\ast _{x}\frac{\mathrm{d}^{2}%
}{\mathrm{d}x^{2}}y\left( x\right) ,\;\;\text{for}\;\;\alpha \in \left(
1,2\right) ,  \label{Dx-1} \\
\mathrm{D}_{x}^{\alpha }y\left( x\right) &=&\frac{\left\vert x\right\vert
^{2-\alpha }\func{sgn}x}{2\Gamma \left( 3-\alpha \right) }\ast _{x}\frac{%
\mathrm{d}^{3}}{\mathrm{d}x^{3}}y\left( x\right) ,\;\;\text{for}\;\;\alpha
\in \left( 2,3\right) ,  \label{Dx-2}
\end{eqnarray}%
with $\ast _{x}$ denoting the convolution in space: $f\left( x\right) \ast
_{x}g\left( x\right) =\int_{%
\mathbb{R}
}f\left( x^{\prime }\right) g\left( x-x^{\prime }\right) \mathrm{d}x^{\prime
},$ $x\in 
\mathbb{R}
.$

The Cauchy problem on the real line $x\in \mathbb{R}$ and $t>0$ is
considered, so the system of governing equations (\ref{eq-motion}), (\ref%
{strejn}), and one of the constitutive equations (\ref{const-eq}), or (\ref%
{UCE-1-5}), or (\ref{UCE-6-8}), or (\ref{nl-Huk}), or (\ref{frac-Eringen})
is subject to initial and boundary conditions: 
\begin{gather}
u(x,0)=u_{0}(x),\;\;\frac{\partial }{\partial t}u(x,0)=v_{0}(x),
\label{ic-u} \\
\sigma (x,0)=0,\;\;\varepsilon (x,0)=0,\;\;\partial _{t}\sigma
(x,0)=0,\;\;\partial _{t}\varepsilon (x,0)=0,  \label{ic-sigma-eps} \\
\lim_{x\rightarrow \pm \infty }u(x,t)=0,\;\;\lim_{x\rightarrow \pm \infty
}\sigma (x,t)=0,  \label{bc}
\end{gather}%
where $u_{0}$ is the initial displacement and $v_{0}$ is the initial
velocity. The initial conditions (\ref{ic-sigma-eps}) are needed for the
hereditary constitutive equations: distributed-order constitutive equation (%
\ref{const-eq}) needs (\ref{ic-sigma-eps})$_{1,2}$ and fractional Burgers
models (\ref{UCE-1-5}), (\ref{UCE-6-8}) require all initial conditions (\ref%
{ic-sigma-eps}), while non-local constitutive models (\ref{nl-Huk}) and (\ref%
{frac-Eringen}) do not need any of the initial conditions (\ref{ic-sigma-eps}%
).

The distributed-order constitutive model (\ref{const-eq}) generalizes
integer and fractional order constitutive models of linear viscoelasticity
having differentiation orders up to the first order, since it reduces to the
linear fractional model%
\begin{equation}
\sum_{i=1}^{n}a_{i}\,{}_{0}\mathrm{D}_{t}^{\alpha _{i}}\sigma
(x,t)=\sum_{j=1}^{m}b_{j}\,{}_{0}\mathrm{D}_{t}^{\beta _{j}}\varepsilon
(x,t),  \label{gen-lin}
\end{equation}%
with model parameters $a_{i},b_{j}>0$ and $\alpha _{i},\beta _{j}\in \left[
0,1\right] ,$ $i=1,\ldots ,n$, $j=1,\ldots ,m,$ if the constitutive
distributions $\phi _{\sigma }$ and $\phi _{\varepsilon }$ in (\ref{const-eq}%
) are chosen as 
\begin{equation*}
\phi _{\sigma }(\alpha )=\sum_{i=1}^{n}a_{i}\,\delta (\alpha -\alpha
_{i}),\;\;\phi _{\varepsilon }(\alpha )=\sum_{j=1}^{m}b_{i}\,\delta (\alpha
-\beta _{j}),
\end{equation*}%
where $\delta $ denotes the Dirac delta distribution. Moreover, the
power-type distributed-order model 
\begin{equation}
\int_{0}^{1}a^{\alpha }\,{}_{0}\mathrm{D}_{t}^{\alpha }\sigma (x,t)\,\mathrm{%
d}\alpha =E\int_{0}^{1}b^{\alpha }\,{}_{0}\mathrm{D}_{t}^{\alpha
}\varepsilon (x,t)\,\mathrm{d}\alpha ,  \label{DOCE}
\end{equation}%
is obtained from (\ref{const-eq}) as the genuine distributed-order model, if
constitutive functions $\phi _{\sigma }$ and $\phi _{\varepsilon }$ in (\ref%
{const-eq}) are chosen as%
\begin{equation*}
\phi _{\sigma }(\alpha )=a^{\alpha },\;\;\phi _{\varepsilon }(\alpha
)=E\,b^{\alpha },
\end{equation*}%
with model parameters $E,a,b>0$ ensuring dimensional homogeneity.

Thermodynamical consistency of linear fractional constitutive equation (\ref%
{gen-lin}) is examined in \cite{AKOZ}, where it is shown that there are four
cases of (\ref{gen-lin}) when the restrictions on model parameters guarantee
its thermodynamical consistency, while power-type distributed-order model (%
\ref{DOCE}) is considered in \cite{a-2003} and revisited in \cite{AKOZ},
where the conditions $E>0$ and $0\leq a\leq b,$ guaranteeing model's
thermodynamical consistency, are obtained. Four cases of thermodynamically
acceptable models corresponding to (\ref{gen-lin}) are given in Appendix \ref%
{LFMS}.

Fractional wave equations, corresponding to the system of governing
equations (\ref{eq-motion}), (\ref{strejn}), and distributed-order
constitutive model (\ref{const-eq}), are considered for the Cauchy problem
in \cite{KOZ19}, generalizing the results of \cite{KOZ10,KOZ11}, where
respectively fractional Zener model and its generalization 
\begin{gather}
\left( 1+a\,{}_{0}\mathrm{D}_{t}^{\alpha }\right) \sigma (x,t)=E\left(
1+b\,{}_{0}\mathrm{D}_{t}^{\alpha }\right) \varepsilon (x,t),\;\;0\leq a\leq
b,\;\alpha \in \left[ 0,1\right] ,  \label{FZM} \\
\sum_{i=1}^{n}a_{i}\,{}_{0}\mathrm{D}_{t}^{\alpha _{i}}\sigma
(x,t)=\sum_{i=1}^{n}b_{i}\,{}_{0}\mathrm{D}_{t}^{\alpha _{i}}\varepsilon
(x,t),\;\;0\leq \alpha _{1}\leq \ldots \leq \alpha _{n}<1,\;\frac{a_{1}}{%
b_{1}}\geq \ldots \geq \frac{a_{n}}{b_{n}}\geq 0,  \notag
\end{gather}%
are considered as special cases of (\ref{gen-lin}). Considering the wave
propagation speed, it is found in \cite{KOZ19} that the finite wave speed,
so as the infinite, is the property of both solid-like and fluid-like
materials. Solid-like and fluid-like materials are differed in the creep
test, representing the deformation response of a material to a sudden but
later on constant stress, where the deformation for first type of materials
is bounded for large time in contrary to the second type of materials that
have unbounded deformation for large time.

Eight thermodynamically consistent fractional Burgers models, formulated in 
\cite{OZ-1}, all describing fluid-like material behavior are divided into
two classes. The first class, represented by (\ref{UCE-1-5}), contains five
models, such that the highest fractional differentiation order of strain is $%
\mu +\eta \in \left[ 1,2\right] ,$ with $\eta \in \left\{ \alpha ,\beta
\right\} ,$ while the highest fractional differentiation order of stress is
either $\gamma \in \left[ 0,1\right] $ in the case of Model I, with $0\leq
\alpha \leq \beta \leq \gamma \leq \mu \leq 1$ and $\eta \in \left\{ \alpha
,\beta ,\gamma \right\} ,$ or $\gamma \in \left[ 1,2\right] $ in the case of
Models II - V, with $0\leq \alpha \leq \beta \leq \mu \leq 1$ and $\left(
\eta ,\gamma \right) \in \left\{ \left( \alpha ,2\alpha \right) ,\left(
\alpha ,\alpha +\beta \right) ,\left( \beta ,\alpha +\beta \right) ,\left(
\beta ,2\beta \right) \right\} $. Note that the fractional differentiation
order of stress is less than the differentiation order of strain regardless
on the interval $\left[ 0,1\right] $ or $\left[ 1,2\right] .$ The second
class, represented by (\ref{UCE-6-8}), contains three models, such that $%
0\leq \alpha \leq \beta \leq 1$ and $\beta +\eta \in \left[ 1,2\right] ,$
with $\eta =\alpha ,$ in the case of Model VI; $\eta =\beta $ in the case of
Model VII; and $\alpha =\eta =\beta ,$ $\bar{a}_{1}=a_{1}+a_{2},$ and $\bar{a%
}_{2}=a_{3}$ in the case of Model VIII. Note that considering the interval $%
\left[ 0,1\right] ,$ the highest fractional differentiation orders of stress
and strain are equal, which also holds true for the orders from interval $%
\left[ 1,2\right] .$ The explicit forms of Models I - VIII, along with
corresponding thermodynamical restrictions, can be found in Appendix \ref%
{FBMS}.

Fractional Burgers wave equation, represented by the governing equations (%
\ref{eq-motion}), (\ref{strejn}), and either (\ref{UCE-1-5}) or (\ref%
{UCE-6-8}), is solved for the Cauchy problem in \cite{OZO}. The wave
propagation speed is found to be infinite for models belonging to the first
class, given by (\ref{UCE-1-5}), contrary to the case of models of the
second class (\ref{UCE-6-8}), that yield finite wave propagation speed.
Moreover, numerical examples indicated that at the wave front there might
exist a jump from finite to a zero value of displacement, obtained as the
fundamental solution of the fractional Burgers equation.

The non-local Hooke law (\ref{nl-Huk}) is introduced in \cite{A-S-09}
through the non-local strain measure and used with the classical Hooke law
as a constitutive equation for modeling wave propagation in non-local media,
while in \cite{AJOPZ} the constitutive equation including both memory and
non-local effects is constructed using fractional Zener model (\ref{FZM})
and non-local Hooke law (\ref{nl-Huk}), further to be used in describing
wave propagation in non-local viscoelastic material. The tools of microlocal
analysis are employed in \cite{HOZ16} to investigate properties of this
memory and non-local type fractional wave equation.

Generalizing the integer-order Eringen stress gradient non-local
constitutive law, the fractional Eringen model (\ref{frac-Eringen}) is
postulated in \cite{CZAS}, where the optimal values of non-locality
parameter and order of fractional differentiation are obtained with respect
to the Born-K\'{a}rm\'{a}n model of lattice dynamics. Further, wave
propagation, as well as propagation of singularities, in non-local material
described by the fractional Eringen model (\ref{frac-Eringen}) is analyzed
in \cite{HOZ18}.

The energy estimates for proving existence and uniqueness of the solution to
three-dimensional wave equation corresponding to material of fractional
Zener type using the Galerkin method are considered in \cite%
{OparnicaSuli,Saedpanah}, while three-dimensional wave equation as a singular kernel integrodifferential equation, with kernel being the relaxation modulus unbounded at the initial time, is analyzed in \cite{Carillo2019}.
 The positivity of Green's functions corresponding to a
three-dimensional integrodifferential{\ }wave equation, that has completely
monotonic relaxation modulus as a kernel, is established in \cite{Ser-1},
while the exponential energy decay of non-linear viscoelastic wave equation
under the potential well is analyzed in \cite{YWangYWang} assuming Dirichlet
boundary conditions.

In the case of {one-dimensional wave equation, written as the
integrodifferential equation including the relaxation modulus assumed as a
wedge continuous function, the solution existence and uniqueness analysis is
performed in \cite{Carillo2019a}, while \cite{Carillo2015} aimed to
underline the similarities between rigid heat conductor having heat flux
relaxation function singular at the origin and viscoelastic material having
relaxation modulus unbounded at the origin. }In \cite{Hanyga2013,Hanyga2019}
one-dimensional wave propagation characteristics, such as wave propagation
speed and wave attenuation, are investigated without and with the Newtonian
viscosity component present in the completely monotonic relaxation modulus.
The extensive overview of wave propagation problems in viscoelastic
materials can be found in \cite{APSZ-2,Holm-book,Mai-10}.

In \cite{Wu} the transient effects, i.e., short-lived seismic wave
propagation through viscoelastic subsurface media, are considered and
asymptotic expansions of the solutions via Buchen-Mainardi algorithm method
introduced in \cite{BuchenMainardi} are obtained. The same method is used in 
\cite{ColombaroGiustiMainardi1} in the case of waves in fractional Maxwell
and Kelvin-Voigt viscoelastic materials. Dispersion, attenuation, wave
fronts, and asymptotic behavior of solution to viscoelastic wave equation
near the wave front are studied in \cite{Han7,Han8,Han6}.

The {survey of acoustic wave equations aiming to describe the frequency
dependent attenuation and scattering of acoustic disturbance propagation
through complex media displaying viscous dissipation is presented in }\cite%
{Cai2018}, while the frequency responses of viscoelastic materials are
reviewed in \cite{Makris}.

The existence and uniqueness of solutions to three-dimensional wave equation
with the Eringen model as a constitutive equation is studied in \cite%
{EvgrafovBellido} and it is found that the problem is in general ill-posed
in the case of smooth kernels and well-posed in the case of singular,
non-smooth kernels. Considering the longitudinal and shear waves propagation
in non-local medium, the influence of geometric non-linearity is
investigated in \cite{MalkhanovErofeevLeontieva}. Combining viscoelastic and
non-locality characteristics of the medium, the wave propagation and wave
decay is studied in \cite{Silling} under the source positioned at the
end of a semi-infinite medium.

\section{Hereditary fractional wave equations expressed through relaxation
modulus and creep compliance}

Relaxation modulus and creep compliance, representing material properties in
stress relaxation and creep tests, are used in order to formulate fractional
wave equation corresponding to the system of governing equations (\ref%
{eq-motion}), (\ref{strejn}), and (\ref{const-eq}), or (\ref{UCE-1-5}), or (%
\ref{UCE-6-8}).

Relaxation modulus $\sigma _{sr}$ (creep compliance $\varepsilon _{cr}$) is
the stress (strain) history function obtained as a response to the strain
(stress) assumed as the Heaviside step function $H.$ According to the
material behavior in stress relaxation and creep tests at the initial
time-instant, one differs the materials having either finite or infinite
glass modulus $\sigma _{sr}^{\left( g\right) }=\sigma _{sr}\left( 0\right) ,$
implying the finite or zero value of the glass compliance $\varepsilon
_{cr}^{\left( g\right) }=\varepsilon _{cr}\left( 0\right) .$ The wave
propagation speed, obtained as%
\begin{equation*}
c=\sqrt{\sigma _{sr}^{\left( g\right) }}=\frac{1}{\sqrt{\varepsilon
_{cr}^{\left( g\right) }}}
\end{equation*}%
in \cite{KOZ19} for the distributed-order constitutive model (\ref{const-eq}%
) and in \cite{OZO} for the fractional Burgers models (\ref{UCE-1-5}) and (%
\ref{UCE-6-8}), is the implication of these material properties. On the
other hand, according to the material behavior in stress relaxation and
creep tests for large time, one differs fluid-like materials, having the
equilibrium compliance $\varepsilon _{cr}^{\left( e\right)
}=\lim_{t\rightarrow \infty }\varepsilon _{cr}\left( t\right) $ infinite and
therefore the equilibrium modulus $\sigma _{sr}^{\left( e\right)
}=\lim_{t\rightarrow \infty }\sigma _{sr}\left( t\right) $ zero, from
solid-like materials, having both equilibrium compliance and equilibrium
modulus finite. The overview of asymptotic properties for viscoelastic
materials described by constitutive models (\ref{const-eq}), (\ref{UCE-1-5}%
), and (\ref{UCE-6-8}) is presented in Table \ref{tbl}.  \begin{table}[h]
 \begin{center}
 \begin{tabular}{|l|c|c|llll|}
 \hline \xrowht{15pt}
 Model & Material type & Wave speed & $\sigma _{sr}^{\left( g\right) }$ & $ \varepsilon_{cr}^{\left(g\right) }$ & $\sigma _{sr}^{\left( e\right) }$ & $\varepsilon_{cr}^{\left(e\right) }$ \\ \hhline{|=|=|=|====|} \xrowht{18pt}
 Power-type & \multirow{3}{*}{\multirowcell{3}{solid-like}} & \multirow{2}{*}{\multirowcell{2}{finite}} & $\frac{Eb}{a}$ & $\frac{a}{Eb}$ & $E$ & $\frac{1}{E}$ \\ \cline{1-1} \cline{4-7} \xrowht{18pt} 
 Case I & & & $\frac{b_{n}}{a_{n}}$ & $\frac{a_{n}}{b_{n}}$ & $\frac{b_{1}}{a_{1}} $ & $\frac{a_{1}}{b_{1}}$ \\ \cline{1-1} \cline{3-7} \xrowht{18pt} 
 Case II & & infinite & $\infty $ & $0$ & $\frac{b_{1}}{a_{1}}$ & $\frac{a_{1}}{b_{1}}$ \\ \hline \xrowht{18pt} 
 Case III & \multirow{4}{*}{\multirowcell{4}{fluid-like}} & finite & $\frac{b_{n}}{a_{n}}$ & $\frac{a_{n}}{b_{n}}$ & $0$ & $\infty $ \\ \cline{1-1} \cline{3-7} \xrowht{18pt} 
 Case IV & & \multirow{2}{*}{\multirowcell{2}{infinite}} & $\infty $ & $0$ & $0$ & $\infty$ \\ \cline{1-1} \cline{4-7} \xrowht{18pt}
 Models I - V & & & $\infty $ & $0$ & $0$ & $\infty$ \\ \cline{1-1} \cline{3-7} \xrowht{18pt}
 Models VI - VIII &  & finite & $\frac{b_{2}}{a_{3}}$ & $\frac{a_{3}}{b_{2}}$ & $0$ & $\infty$ \\ 
 \hline
 \end{tabular}
 \end{center}
 \caption{Summary of models' properties.}
 \label{tbl}
 \end{table}

In order to express the constitutive equations (\ref{const-eq}), (\ref%
{UCE-1-5}), and (\ref{UCE-6-8}) either in terms of the relaxation modulus,
or in terms of the creep compliance, the Laplace transform with respect to
time%
\begin{equation*}
\tilde{f}\left( s\right) =\mathcal{L}\left[ f\left( t\right) \right] \left(
s\right) =\int_{0}^{\infty }f\left( t\right) \mathrm{e}^{-st}\,\mathrm{d}%
t,\;\;\func{Re}s>0,
\end{equation*}%
is applied to (\ref{const-eq}), (\ref{UCE-1-5}), and (\ref{UCE-6-8}), so
that 
\begin{equation}
\Phi _{\sigma }(s)\tilde{\sigma}\left( x,s\right) =\Phi _{\varepsilon }(s)%
\tilde{\varepsilon}\left( x,s\right) ,\;\;\func{Re}s>0,  \label{CEs-LT}
\end{equation}%
is obtained assuming zero initial conditions (\ref{ic-sigma-eps}), with%
\begin{equation}
\Phi _{\sigma }(s)=\int_{0}^{1}\phi _{\sigma }(\alpha )s^{\alpha }\,\mathrm{d%
}\alpha ,\;\;\Phi _{\varepsilon }(s)=\int_{0}^{1}\phi _{\varepsilon }(\alpha
)s^{\alpha }\,\mathrm{d}\alpha ,  \label{fiovi}
\end{equation}%
in the case distributed-order constitutive model (\ref{const-eq}), reducing
to%
\begin{equation}
\Phi _{\sigma }(s)=\sum_{i=1}^{n}a_{i}\,s^{\alpha _{i}},\;\;\Phi
_{\varepsilon }(s)=\sum_{j=1}^{m}b_{j}\,s^{\beta _{j}},\;\;\text{and}%
\;\;\Phi _{\sigma }(s)=\frac{as-1}{\ln \left( as\right) },\;\;\Phi
_{\varepsilon }(s)=E\frac{bs-1}{\ln \left( bs\right) },  \label{fiovi-lin}
\end{equation}%
for linear fractional constitutive equation (\ref{gen-lin}) and power-type
distributed-order model (\ref{DOCE}), respectively, as well as with%
\begin{gather}
\Phi _{\sigma }(s)=1+a_{1}s^{\alpha }+a_{2}\,s^{\beta }+a_{3}\,s^{\gamma
},\;\;\Phi _{\varepsilon }(s)=b_{1}\,s^{\mu }+b_{2}\,s^{\mu +\eta },
\label{Burgers1-fiovi} \\
\Phi _{\sigma }(s)=1+a_{1}s^{\alpha }+a_{2}\,s^{\beta }+a_{3}\,s^{\beta
+\eta },\;\;\Phi _{\varepsilon }(s)=b_{1}\,s^{\beta }+b_{2}\,s^{\beta +\eta
},  \label{Burgers2-fiovi}
\end{gather}%
in the case of fractional Burgers model of the first, respectively second
class, given by (\ref{UCE-1-5}) and (\ref{UCE-6-8}).

The Laplace transform of relaxation modulus and creep compliance%
\begin{equation}
\tilde{\sigma}_{sr}\left( s\right) =\frac{1}{s}\frac{\Phi _{\varepsilon }(s)%
}{\Phi _{\sigma }(s)}\;\;\text{and}\;\;\tilde{\varepsilon}_{cr}\left(
s\right) =\frac{1}{s}\frac{\Phi _{\sigma }(s)}{\Phi _{\varepsilon }(s)}
\label{sr-cr}
\end{equation}%
are respectively obtained by using the Laplace transform of constitutive
equation (\ref{CEs-LT}) for $\tilde{\varepsilon}\left( x,s\right) =\mathcal{L%
}\left[ H\left( t\right) \right] \left( s\right) =\frac{1}{s}$ and $\tilde{%
\sigma}\left( x,s\right) =\mathcal{L}\left[ H\left( t\right) \right] \left(
s\right) =\frac{1}{s},$ so that (\ref{sr-cr}) used in (\ref{CEs-LT}) yielded
the Laplace transform of constitutive equation (\ref{CEs-LT}) expressed
either in terms of relaxation modulus, or in terms of creep compliance as%
\begin{equation}
\frac{1}{s}\tilde{\sigma}\left( x,s\right) =\tilde{\sigma}_{sr}\left(
s\right) \tilde{\varepsilon}\left( x,s\right) \;\;\text{or}\;\;\frac{1}{s}%
\tilde{\varepsilon}\left( x,s\right) =\tilde{\varepsilon}_{cr}\left(
s\right) \tilde{\sigma}\left( x,s\right) ,  \label{CEs-LT-1}
\end{equation}%
providing six equivalent forms of the hereditary constitutive equation:
three expressed in terms of relaxation modulus 
\begin{gather}
\int_{0}^{t}\sigma \left( x,t^{\prime }\right) \mathrm{d}t^{\prime }=\sigma
_{sr}\left( t\right) \ast _{t}\varepsilon \left( x,t\right) ,
\label{int-sigma-eps-1} \\
\sigma \left( x,t\right) =\sigma _{sr}^{\left( g\right) }\varepsilon \left(
x,t\right) +\dot{\sigma}_{sr}\left( t\right) \ast _{t}\varepsilon \left(
x,t\right) ,  \label{sigma-konv} \\
\sigma \left( x,t\right) =\sigma _{sr}\left( t\right) \ast _{t}\partial
_{t}\varepsilon \left( x,t\right) ,  \label{sigma-konv-1}
\end{gather}%
obtained by the Laplace transform inversion in (\ref{CEs-LT-1})$_{1}$ and
three expressed in terms of creep compliance%
\begin{gather}
\int_{0}^{t}\varepsilon \left( x,t^{\prime }\right) \mathrm{d}t^{\prime
}=\varepsilon _{cr}\left( t\right) \ast _{t}\sigma \left( x,t\right) ,
\label{int-sigma-eps-2} \\
\varepsilon \left( x,t\right) =\varepsilon _{cr}^{\left( g\right) }\sigma
\left( x,t\right) +\dot{\varepsilon}_{cr}\left( t\right) \ast _{t}\sigma
\left( x,t\right) ,  \label{eps-konv} \\
\varepsilon \left( x,t\right) =\varepsilon _{cr}\left( t\right) \ast
_{t}\partial _{t}\sigma \left( x,t\right) ,  \label{eps-konv-1}
\end{gather}%
obtained by the Laplace transform inversion in (\ref{CEs-LT-1})$_{2},$ with $%
\dot{f}\left( t\right) =\frac{\mathrm{d}}{\mathrm{d}t}f\left( t\right) $ and
by using $\frac{\mathrm{d}}{\mathrm{d}t}\left( f\left( t\right) \ast
_{t}g\left( t\right) \right) =f\left( 0\right) g\left( t\right) +\dot{f}%
\left( t\right) \ast _{t}g\left( t\right) ,$ along with the initial
conditions on stress and strain (\ref{ic-sigma-eps}).

Therefore, the equivalent forms of hereditary fractional wave equation
expressed in terms of relaxation modulus%
\begin{gather}
\rho \,\partial _{t}u\left( x,t\right) =\rho \,v_{0}\left( x\right) +\sigma
_{sr}\left( t\right) \ast _{t}\partial _{xx}u\left( x,t\right) ,  \notag \\
\rho \,\partial _{tt}u\left( x,t\right) =\sigma _{sr}^{\left( g\right)
}\,\partial _{xx}u\left( x,t\right) +\dot{\sigma}_{sr}\left( t\right) \ast
_{t}\partial _{xx}u\left( x,t\right) ,  \label{FWE-sigma-g} \\
\rho \,\partial _{tt}u\left( x,t\right) =\sigma _{sr}\left( t\right) \ast
_{t}\partial _{txx}u\left( x,t\right) ,  \label{FWE-sigma}
\end{gather}%
are respectively obtained by differentiation of (\ref{int-sigma-eps-1}), (%
\ref{sigma-konv}), and (\ref{sigma-konv-1}) with respect to the spatial
coordinate and by the subsequent use of equation of motion (\ref{eq-motion})
and strain (\ref{strejn}) in such obtained expressions, including the
initial condition (\ref{ic-u})$_{2},$ while the equivalent forms of
hereditary fractional wave equation expressed in terms of creep compliance%
\begin{gather}
\rho \,\varepsilon _{cr}\left( t\right) \ast _{t}\partial _{tt}u\left(
x,t\right) =\int_{0}^{t}\partial _{xx}u\left( x,t^{\prime }\right) \mathrm{d}%
t^{\prime },  \notag \\
\rho \varepsilon _{cr}^{\left( g\right) }\,\partial _{tt}u\left( x,t\right)
+\rho \,\dot{\varepsilon}_{cr}\left( t\right) \ast _{t}\partial _{tt}u\left(
x,t\right) =\partial _{xx}u\left( x,t\right) ,  \label{FWE-epsilon-g} \\
\rho \,\varepsilon _{cr}\left( t\right) \ast _{t}\partial _{tt{}t}u\left(
x,t\right) =\partial _{xx}u\left( x,t\right) ,  \notag
\end{gather}%
are respectively obtained by differentiation of (\ref{int-sigma-eps-2}), (%
\ref{eps-konv}), and (\ref{eps-konv-1}) with respect to the spatial
coordinate and by the subsequent use of equation of motion (\ref{eq-motion})
and strain (\ref{strejn}) in such obtained expressions.

\section{Relaxation modulus and creep compliance}

Starting from the distributed-order viscoelastic model (\ref{const-eq})
having differentiation order below the first order, the conditions for
relaxation modulus to be completely monotonic and simultaneously creep
compliance to be Bernstein function are derived by the means of Laplace
transform method. It is shown that these conditions for relaxation modulus
and creep compliance in cases of linear fractional models (\ref{gen-lin})
and power-type distributed-order model (\ref{DOCE}) are equivalent to the
thermodynamical requirements implying four thermodynamically acceptable
cases of linear fractional models (\ref{gen-lin}), listed in Appendix \ref%
{LFMS}, and the power-type model (\ref{DOCE}), with $E>0$ and $0\leq a\leq
b. $ These properties of creep compliance and relaxation modulus are proved
to be of crucial importance in establishing dissipativity of the hereditary
fractional wave equation. Recall, completely monotonic function is a
positive, monotonically decreasing convex function, or more precisely
function $f$ satisfying $\left( -1\right) ^{n}f^{\left( n\right) }\left(
t\right) \geq 0,$ $n\in 
\mathbb{N}
_{0},$ while Bernstein function is a positive, monotonically increasing,
concave function, or more precisely non-negative function having its first
derivative completely monotonic.

The responses in creep and stress relaxation tests of thermodynamically
consistent fractional Burgers models (\ref{UCE-1-5}) and (\ref{UCE-6-8}) are
examined in \cite{OZ-2}, where it is found that the requirements for
relaxation modulus to be completely monotonic and creep compliance to be
Bernstein function are more restrictive than the thermodynamical
requirements. Conditions guaranteeing the thermodynamical consistency of
fractional Burgers models and narrower conditions guaranteeing monotonicity
properties of relaxation modulus and creep compliance are given in Appendix %
\ref{FBMS}.

The relaxation modulus, corresponding to the distributed-order viscoelastic
model (\ref{const-eq}), takes the form%
\begin{eqnarray}
\sigma _{sr}\left( t\right) &=&\sigma _{sr}^{\left( e\right) }+\frac{1}{\pi }%
\int_{0}^{\infty }\frac{K\left( \rho \right) }{\left\vert \Phi _{\sigma
}\left( \rho \mathrm{e}^{\mathrm{i}\pi }\right) \right\vert ^{2}}\frac{%
\mathrm{e}^{-\rho t}}{\rho }\mathrm{d}\rho ,\;\;\text{with}
\label{sigma-sr-eq} \\
\sigma _{sr}^{\left( e\right) } &=&\lim_{t\rightarrow \infty }\sigma
_{sr}\left( t\right) =\lim_{s\rightarrow 0}\left( s\tilde{\sigma}_{sr}\left(
s\right) \right) =\lim_{s\rightarrow 0}\frac{\Phi _{\varepsilon }(s)}{\Phi
_{\sigma }(s)},\;\;\text{and}  \label{sigma-sr-e} \\
K\left( \rho \right) &=&\func{Re}\Phi _{\sigma }\left( \rho \mathrm{e}^{%
\mathrm{i}\pi }\right) \func{Im}\Phi _{\varepsilon }\left( \rho \mathrm{e}^{%
\mathrm{i}\pi }\right) -\func{Im}\Phi _{\sigma }\left( \rho \mathrm{e}^{%
\mathrm{i}\pi }\right) \func{Re}\Phi _{\varepsilon }\left( \rho \mathrm{e}^{%
\mathrm{i}\pi }\right) ,  \label{K}
\end{eqnarray}%
where functions $\Phi _{\sigma }$ and $\Phi _{\varepsilon }$ are defined by (%
\ref{fiovi}), while the creep compliance may be represented either by%
\begin{eqnarray}
\varepsilon _{cr}\left( t\right) &=&\varepsilon _{cr}^{\left( e\right) }-%
\frac{1}{\pi }\int_{0}^{\infty }\frac{K\left( \rho \right) }{\left\vert \Phi
_{\varepsilon }\left( \rho \mathrm{e}^{\mathrm{i}\pi }\right) \right\vert
^{2}}\frac{\mathrm{e}^{-\rho t}}{\rho }\mathrm{d}\rho ,\;\text{with}
\label{eps-cr-eq} \\
\varepsilon _{cr}^{\left( e\right) } &=&\lim_{t\rightarrow \infty
}\varepsilon _{cr}\left( t\right) =\lim_{s\rightarrow 0}\left( s\tilde{%
\varepsilon}_{cr}\left( s\right) \right) =\lim_{s\rightarrow 0}\frac{\Phi
_{\sigma }(s)}{\Phi _{\varepsilon }(s)},  \label{eps-cr-e}
\end{eqnarray}%
for solid-like materials, or by%
\begin{equation}
\varepsilon _{cr}\left( t\right) =\frac{1}{\pi }\int_{0}^{\infty }\frac{%
K\left( \rho \right) }{\left\vert \Phi _{\varepsilon }\left( \rho \mathrm{e}%
^{\mathrm{i}\pi }\right) \right\vert ^{2}}\frac{1-\mathrm{e}^{-\rho t}}{\rho 
}\mathrm{d}\rho ,  \label{eps-cr}
\end{equation}%
for fluid-like materials, where function $K$ is given by (\ref{K}). The
calculation of relaxation modulus (\ref{sigma-sr-eq}) and creep compliances (%
\ref{eps-cr-eq}) and (\ref{eps-cr}) is performed in Appendix \ref{sr-cr-calc}%
.

The equilibrium modulus $\sigma _{sr}^{\left( e\right) }$ has either zero or
finite non-zero value, as seen from Table \ref{tbl}, hence the relaxation
modulus (\ref{sigma-sr-eq}) has the same form regardless of the material
type, while the equilibrium compliance $\varepsilon _{cr}^{\left( e\right) }$
has either finite value for solid-like materials (power-type
distributed-order constitutive equation (\ref{DOCE}) and Cases I (\ref{Case
1}) and II (\ref{Case 2})), or infinite value for fluid-like materials
(Cases III (\ref{Case 3}) and IV (\ref{Case 4})), as summarized in Table \ref%
{tbl}, implying the need for expressing the creep compliance either in form (%
\ref{eps-cr-eq}), or in the form (\ref{eps-cr}).

The function $K,$ calculated by (\ref{K}), for linear fractional models (\ref%
{gen-lin}) and power-type distributed-order model (\ref{DOCE}) takes the
respective forms%
\begin{eqnarray}
K\left( \rho \right) &=&-\sum_{i=1}^{n}\sum_{j=1}^{m}a_{i}b_{j}\,\rho
^{\alpha _{i}+\beta _{j}}\sin \frac{\left( \alpha _{i}-\beta _{j}\right) \pi 
}{2}\;\;\text{and}  \label{K-gen-lin} \\
K\left( \rho \right) &=&E\pi \frac{a\rho +1}{\left\vert \ln \left( a\rho
\right) +\mathrm{i}\pi \right\vert ^{2}}\frac{b\rho +1}{\left\vert \ln
\left( b\rho \right) +\mathrm{i}\pi \right\vert ^{2}}\ln \frac{b}{a},
\label{K-PTDO}
\end{eqnarray}%
obtained by substitution $s=\rho \mathrm{e}^{\mathrm{i}\pi }$ in (\ref%
{fiovi-lin}). By requiring non-negativity of function $K,$ the conditions on
model parameters guaranteeing that the relaxation modulus (\ref{sigma-sr-eq}%
) is completely monotonic, while creep compliances (\ref{eps-cr-eq}) and (%
\ref{eps-cr}) are Bernstein functions are derived, since the non-negativity
of $K$ implies%
\begin{equation*}
\sigma _{sr}\left( t\right) \geq 0\;\;\text{and}\;\;\left( -1\right) ^{k}%
\frac{\mathrm{d}^{k}}{\mathrm{d}t^{k}}\sigma _{sr}\left( t\right) =\frac{1}{%
\pi }\int_{0}^{\infty }\frac{K\left( \rho \right) }{\left\vert \Phi _{\sigma
}\left( \rho \mathrm{e}^{\mathrm{i}\pi }\right) \right\vert ^{2}}\rho ^{k-1}%
\mathrm{e}^{-\rho t}\mathrm{d}\rho \geq 0,\;\;k\in 
\mathbb{N}
,\;t>0,
\end{equation*}%
for relaxation modulus (\ref{sigma-sr-eq}) and%
\begin{equation*}
\varepsilon _{cr}\left( t\right) \geq 0\;\;\text{and}\;\;\left( -1\right)
^{k}\frac{\mathrm{d}^{k}}{\mathrm{d}t^{k}}\dot{\varepsilon}_{cr}\left(
t\right) =\frac{1}{\pi }\int_{0}^{\infty }\frac{K\left( \rho \right) }{%
\left\vert \Phi _{\varepsilon }\left( \rho \mathrm{e}^{\mathrm{i}\pi
}\right) \right\vert ^{2}}\rho ^{k}\mathrm{e}^{-\rho t}\mathrm{d}\rho \geq
0,\;\;k\in 
\mathbb{N}
_{0},\;t>0,
\end{equation*}%
with $\dot{\varepsilon}_{cr}\left( t\right) =\frac{\mathrm{d}}{\mathrm{d}t}%
\varepsilon _{cr}\left( t\right) ,$ for the creep compliances (\ref%
{eps-cr-eq}) and (\ref{eps-cr}). Note that $\dot{\varepsilon}_{cr}\left(
t\right) \geq 0$ in the case of (\ref{eps-cr-eq}) implies that the creep
compliance $\varepsilon _{cr}\left( t\right) $ monotonically increases from $%
\varepsilon _{cr}^{\left( g\right) }=\lim_{s\rightarrow \infty }\frac{\Phi
_{\sigma }(s)}{\Phi _{\varepsilon }(s)}$ to $\varepsilon _{cr}^{\left(
e\right) }=\lim_{s\rightarrow 0}\frac{\Phi _{\sigma }(s)}{\Phi _{\varepsilon
}(s)}$ for $t>0,$ thus being a non-negative function since $\Phi _{\sigma }$
and $\Phi _{\varepsilon }$ are non-negative functions.

By requiring non-negativity of function $K,$ given by (\ref{K-gen-lin}), one
reobtains all four cases of linear fractional model (\ref{gen-lin}), listed
in Appendix \ref{LFMS} along with the explicit forms of corresponding
function $K,$ since by (\ref{K-gen-lin}) function $K$ is up to the
multiplication by the positive function exactly the loss modulus, see \cite[%
Eq. (2.9)]{AKOZ}, whose non-negativity requirement for all (positive)
frequencies yielded four thermodynamically consistent classes of linear
fractional models (\ref{gen-lin}). In the case of function $K$ given by (\ref%
{K-PTDO}), the thermodynamical requirements $E>0$ and $0\leq a\leq b$
guarantee non-negativity of function $K.$

The relaxation modulus (\ref{sigma-sr-eq}) and creep compliances (\ref%
{eps-cr-eq}) and (\ref{eps-cr}) are obtained in Appendix \ref{sr-cr-calc}
under the following assumptions.

\begin{itemize}
\item[$\left( A1\right) $] Functions $\Phi _{\sigma }$ and $\Phi
_{\varepsilon },$ given by (\ref{fiovi}), except for $s=0,$ have no other
branching points and also $\Phi _{\sigma }(s)\neq 0$ and $\Phi _{\varepsilon
}(s)\neq 0$ for $s\in 
\mathbb{C}
,$ implying the nonexistence of poles of functions $\frac{\Phi _{\sigma }(s)%
}{\Phi _{\varepsilon }(s)}$ and $\frac{\Phi _{\varepsilon }(s)}{\Phi
_{\sigma }(s)}$ in the complex plane.

\item[$\left( A2\right) $] In order to obtain the relaxation modulus (\ref%
{sigma-sr-eq}), functions $\Phi _{\sigma }$ and $\Phi _{\varepsilon }$ (\ref%
{fiovi}) should satisfy%
\begin{equation*}
\frac{1}{R}\left\vert \frac{\Phi _{\varepsilon }\left( R\mathrm{e}^{\mathrm{i%
}\frac{\pi }{2}}\right) }{\Phi _{\sigma }(R\mathrm{e}^{\mathrm{i}\frac{\pi }{%
2}})}\right\vert \rightarrow 0\;\;\text{and therefore}\;\;\left\vert \frac{%
\Phi _{\varepsilon }(R\mathrm{e}^{\mathrm{i}\varphi })}{\Phi _{\sigma }(R%
\mathrm{e}^{\mathrm{i}\varphi })}\right\vert \mathrm{e}^{Rt\cos \varphi
}\rightarrow 0,\;\;\text{as}\;\;R\rightarrow \infty ,
\end{equation*}%
for $\varphi \in \left( -\pi ,-\frac{\pi }{2}\right) \cup \left( \frac{\pi }{%
2},\pi \right) .$

\item[$\left( A3\right) $] In order to obtain the creep compliance (\ref%
{eps-cr-eq}), functions $\Phi _{\sigma }$ and $\Phi _{\varepsilon }$ (\ref%
{fiovi}) should satisfy%
\begin{equation*}
\frac{1}{R}\left\vert \frac{\Phi _{\sigma }\left( R\mathrm{e}^{\mathrm{i}%
\frac{\pi }{2}}\right) }{\Phi _{\varepsilon }(R\mathrm{e}^{\mathrm{i}\frac{%
\pi }{2}})}\right\vert \rightarrow 0\;\;\text{and therefore}\;\;\left\vert 
\frac{\Phi _{\sigma }(R\mathrm{e}^{\mathrm{i}\varphi })}{\Phi _{\varepsilon
}(R\mathrm{e}^{\mathrm{i}\varphi })}\right\vert \mathrm{e}^{Rt\cos \varphi
}\rightarrow 0,\;\;\text{as}\;\;R\rightarrow \infty ,
\end{equation*}%
for $\varphi \in \left( -\pi ,-\frac{\pi }{2}\right) \cup \left( \frac{\pi }{%
2},\pi \right) .$

\item[$\left( A4\right) $] In order to obtain the creep compliance (\ref%
{eps-cr}), functions $\Phi _{\sigma }$ and $\Phi _{\varepsilon }$ (\ref%
{fiovi}) should satisfy%
\begin{equation*}
\left\vert \frac{\Phi _{\sigma }\left( R\mathrm{e}^{\mathrm{i}\frac{\pi }{2}%
}\right) }{\Phi _{\varepsilon }(R\mathrm{e}^{\mathrm{i}\frac{\pi }{2}})}%
\right\vert \rightarrow 0\;\;\text{and therefore}\;\;R\left\vert \frac{\Phi
_{\sigma }(R\mathrm{e}^{\mathrm{i}\varphi })}{\Phi _{\varepsilon }(R\mathrm{e%
}^{\mathrm{i}\varphi })}\right\vert \mathrm{e}^{Rt\cos \varphi }\rightarrow
0,\;\;\text{or}\;\;p_{0}=0,\;\;\text{as}\;\;R\rightarrow \infty ,
\end{equation*}%
for $\varphi \in \left( -\pi ,-\frac{\pi }{2}\right) \cup \left( \frac{\pi }{%
2},\pi \right) ,$ as well as%
\begin{equation*}
r\left\vert \frac{\Phi _{\sigma }(r\mathrm{e}^{\mathrm{i}\varphi })}{\Phi
_{\varepsilon }(r\mathrm{e}^{\mathrm{i}\varphi })}\right\vert \rightarrow
0,\;\;\text{as}\;\;r\rightarrow 0,
\end{equation*}%
for $\varphi \in \left( -\pi ,\pi \right) .$
\end{itemize}

Assumption $\left( A1\right) $ is satisfied for linear fractional models (%
\ref{gen-lin}) as well as for the power-type model (\ref{DOCE}), due to the
fractional differentiation orders belonging to the interval between zero and
one. For thermodynamically acceptable cases of linear fractional models (\ref%
{gen-lin}), listed in Appendix \ref{LFMS}, and for the power-type model (\ref%
{DOCE}) assumption $\left( A2\right) $ is satisfied, since either $%
\left\vert \frac{\Phi _{\varepsilon }(R\mathrm{e}^{\mathrm{i}\varphi })}{%
\Phi _{\sigma }(R\mathrm{e}^{\mathrm{i}\varphi })}\right\vert \sim C$ or $%
\left\vert \frac{\Phi _{\varepsilon }(R\mathrm{e}^{\mathrm{i}\varphi })}{%
\Phi _{\sigma }(R\mathrm{e}^{\mathrm{i}\varphi })}\right\vert \sim \frac{C}{%
R^{\delta }},$ as $R\rightarrow \infty ,$ with $C$ being constant and $%
\delta \in \left( 0,1\right) ,$ see Table \ref{tbl-1}. As already
anticipated, constitutive equations corresponding to the solid-like
materials (power-type distributed-order constitutive equation (\ref{DOCE})
and Cases I (\ref{Case 1}) and II (\ref{Case 2})) satisfy assumption $\left(
A3\right) ,$ while constitutive equations corresponding to the fluid-like
materials (Cases III (\ref{Case 3}) and IV (\ref{Case 4})) satisfy
assumption $\left( A4\right) ,$ see Table \ref{tbl-1}.  \begin{table}[h]
 \begin{center}
 \begin{tabular}{|l|l|l|l|} 
 \hline \xrowht{18pt} 
 Model & 
 $\left\vert \frac{\Phi _{\sigma }(R\mathrm{e}^{\mathrm{i}\varphi })}{\Phi_{\varepsilon }(R\mathrm{e}^{\mathrm{i}\varphi })}\right\vert $ as $R\rightarrow \infty $ & 
 $\left\vert \frac{\Phi _{\varepsilon}(R\mathrm{e}^{\mathrm{i}\varphi })}{\Phi _{\sigma}(R\mathrm{e}^{\mathrm{i}\varphi })}\right\vert $ as $R\rightarrow \infty $ &
 $\left\vert \frac{\Phi _{\sigma }(r\mathrm{e}^{\mathrm{i}\varphi })}{\Phi_{\varepsilon }(r\mathrm{e}^{\mathrm{i}\varphi })}\right\vert $ as $r\rightarrow 0$ \\ \hhline{|=|=|=|=|} \xrowht{18pt} 
 Power-type & $\sim \frac{a}{Eb}$ & $\sim \frac{Eb}{a}$ & $\sim \frac{1}{E}$ \\ \hline \xrowht{18pt}
 Case I & $\sim \frac{a_{n}}{b_{n}}$ & $\sim \frac{b_{n}}{a_{n}}$ & $\sim\frac{a_{1}}{b_{1}}$ \\ \hline \xrowht{18pt}
 Case II & $\sim \frac{a_{n}}{b_{m}}\frac{1}{R^{\beta _{m}-\alpha _{n}}}$ & $\sim \frac{b_{m}}{a_{n}}R^{\beta _{m}-\alpha _{n}}$ & $\sim \frac{a_{1}}{b_{1}}$ \\ \hline \xrowht{18pt} 
 Case III & $\sim \frac{a_{n}}{b_{n}}$ & $\sim \frac{b_{n}}{a_{n}}$ & $\sim \frac{a_{1}}{b_{m+1}}\frac{1}{r^{\alpha_{m+1}-\alpha _{1}}}$ \\ \hline \xrowht{18pt}
 Case IV & $\sim \frac{a_{n}}{b_{m}} \frac{1}{R^{\beta _{m}-\alpha _{n}}}$ & $\sim \frac{b_{m}}{a_{n}}R^{\beta_{m}-\alpha _{n}}$ & $\sim \frac{a_{1}}{b_{1}}\frac{1}{r^{\beta _{1}-\alpha _{1}}}$ \\ \hline
 \end{tabular}
 \end{center}
 \caption{Asymptotics of models originating from the distributed-order model (\ref{const-eq}).}
 \label{tbl-1}
 \end{table}

\section{Energy dissipation for hereditary materials}

A priori energy estimates stating that the kinetic energy at arbitrary
time-instant is less than the initial kinetic energy are derived in order to
show the dissipativity of the hereditary fractional wave equations. The
material properties at initial time-instant, differing the materials with
finite and infinite wave propagation speed, prove to have a decisive role in
choosing the form of fractional wave equation and the form of energy
estimates as well. In proving dissipativity properties of the hereditary
fractional wave equations, the key point is that relaxation modulus is
completely monotonic function. Similarly, energy estimate involving the
creep compliance is based on the fact that the creep compliance is Bernstein
function.

\subsection{Materials having finite glass modulus}

The energy estimate for fractional wave equation expressed in terms of
relaxation modulus (\ref{FWE-sigma-g}) correspond to materials that have
finite glass modulus and thus finite wave speed as well, i.e., materials
described by the power-type distributed-order model (\ref{DOCE}), Case I (%
\ref{Case 1}), and Case III (\ref{Case 3}) of the linear constitutive model (%
\ref{gen-lin}), as well as materials described by the fractional Burgers
models VI - VIII (\ref{Model 6}), (\ref{Model 7}), (\ref{Model 8}).

Namely, by multiplying the fractional wave equation (\ref{FWE-sigma-g}) by $%
\partial _{t}u$ and by subsequent integration with respect to the spatial
coordinate along the whole domain $%
\mathbb{R}
$ and with respect to time over interval $\left[ 0,t\right] ,$ where $t>0$
is the arbitrary time-instant, one has%
\begin{equation}
\frac{1}{2}\rho \left\Vert \partial _{t}u\left( \cdot ,t\right) \right\Vert
_{L^{2}\left( 
\mathbb{R}
\right) }^{2}+\frac{1}{2}\sigma _{sr}^{\left( g\right) }\left\Vert \partial
_{x}u\left( \cdot ,t\right) \right\Vert _{L^{2}\left( 
\mathbb{R}
\right) }^{2}=\frac{1}{2}\rho \left\Vert v_{0}\left( \cdot \right)
\right\Vert _{L^{2}\left( 
\mathbb{R}
\right) }^{2}+\int_{0}^{t}\!\!\int_{\mathbb{R}}\left( \dot{\sigma}%
_{sr}\left( t^{\prime }\right) \ast _{t^{\prime }}\partial _{xx}u\left(
x,t^{\prime }\right) \right) \,\partial _{t^{\prime }}u\left( x,t^{\prime
}\right) \,\mathrm{d}x\,\mathrm{d}t^{\prime },  \label{ee-sigma-g-skoro}
\end{equation}%
where the change of kinetic energy (per unit square) of a viscoelastic
(infinite) body is obtained as%
\begin{eqnarray}
\rho \int_{0}^{t}\!\!\int_{\mathbb{R}}\partial _{t^{\prime }t^{\prime
}}u\left( x,t^{\prime }\right) \,\partial _{t^{\prime }}u\left( x,t^{\prime
}\right) \,\mathrm{d}x\,\mathrm{d}t^{\prime } &=&\frac{1}{2}\rho
\int_{0}^{t}\!\!\int_{\mathbb{R}}\partial _{t^{\prime }}\left( \partial
_{t^{\prime }}u\left( x,t^{\prime }\right) \right) ^{2}\,\mathrm{d}x\,%
\mathrm{d}t^{\prime }=\frac{1}{2}\rho \int_{0}^{t}\partial _{t^{\prime
}}\left\Vert \partial _{t^{\prime }}u\left( \cdot ,t^{\prime }\right)
\right\Vert _{L^{2}\left( 
\mathbb{R}
\right) }^{2}\,\mathrm{d}t^{\prime }  \notag \\
&=&\frac{1}{2}\rho \left\Vert \partial _{t}u\left( \cdot ,t\right)
\right\Vert _{L^{2}\left( 
\mathbb{R}
\right) }^{2}-\frac{1}{2}\rho \left\Vert v_{0}\left( \cdot \right)
\right\Vert _{L^{2}\left( 
\mathbb{R}
\right) }^{2},  \label{kin-en}
\end{eqnarray}%
using the initial condition (\ref{ic-u})$_{2},$ while the potential energy
(per unit square) of a viscoelastic (infinite) body follows from 
\begin{eqnarray}
\int_{0}^{t}\!\!\int_{\mathbb{R}}\partial _{xx}u\left( x,t^{\prime }\right)
\,\partial _{t^{\prime }}u\left( x,t^{\prime }\right) \,\mathrm{d}x\,\mathrm{%
d}t^{\prime } &=&\int_{0}^{t}\left( \left[ \partial _{x}u\left( x,t^{\prime
}\right) \,\partial _{t^{\prime }}u\left( x,t^{\prime }\right) \right]
_{x\rightarrow -\infty }^{x\rightarrow \infty }-\int_{\mathbb{R}}\partial
_{x}u\left( x,t^{\prime }\right) \,\partial _{xt^{\prime }}u\left(
x,t^{\prime }\right) \,\mathrm{d}x\right) \,\mathrm{d}t^{\prime }  \notag \\
&=&-\frac{1}{2}\int_{0}^{t}\!\!\int_{\mathbb{R}}\partial _{t^{\prime
}}\left( \partial _{x}u\left( x,t^{\prime }\right) \right) ^{2}\,\mathrm{d}%
x\,\mathrm{d}t^{\prime }=-\frac{1}{2}\int_{0}^{t}\partial _{t^{\prime
}}\left\Vert \partial _{x}u\left( \cdot ,t^{\prime }\right) \right\Vert
_{L^{2}\left( 
\mathbb{R}
\right) }^{2}\,\mathrm{d}t^{\prime }  \notag \\
&=&-\frac{1}{2}\left\Vert \partial _{x}u\left( \cdot ,t\right) \right\Vert
_{L^{2}\left( 
\mathbb{R}
\right) }^{2}+\frac{1}{2}\left\Vert \varepsilon \left( x,0\right)
\right\Vert _{L^{2}\left( 
\mathbb{R}
\right) }^{2}=-\frac{1}{2}\left\Vert \partial _{x}u\left( \cdot ,t\right)
\right\Vert _{L^{2}\left( 
\mathbb{R}
\right) }^{2},  \label{pot-en}
\end{eqnarray}%
using the initial condition (\ref{ic-sigma-eps})$_{2}$ and integration by
parts along with the boundary conditions (\ref{bc})$_{2}$ combined with the
constitutive equation (\ref{sigma-konv}) and strain (\ref{strejn}) yielding $%
\lim_{x\rightarrow \pm \infty }\partial _{x}u\left( x,t\right) =0.$

The last term on the right-hand-side of (\ref{ee-sigma-g-skoro}) is
transformed as%
\begin{eqnarray*}
&&\int_{0}^{t}\!\!\int_{\mathbb{R}}\left( \dot{\sigma}_{sr}\left( t^{\prime
}\right) \ast _{t^{\prime }}\partial _{xx}u\left( x,t^{\prime }\right)
\right) \,\partial _{t^{\prime }}u\left( x,t^{\prime }\right) \,\mathrm{d}x\,%
\mathrm{d}t^{\prime } \\
&&\qquad =\int_{0}^{t}\!\!\int_{\mathbb{R}}\partial _{x}\left( \dot{\sigma}%
_{sr}\left( t^{\prime }\right) \ast _{t^{\prime }}\partial _{x}u\left(
x,t^{\prime }\right) \right) \,\partial _{t^{\prime }}u\left( x,t^{\prime
}\right) \,\mathrm{d}x\,\mathrm{d}t^{\prime } \\
&&\qquad =\int_{0}^{t}\left( \left[ \left( \dot{\sigma}_{sr}\left( t^{\prime
}\right) \ast _{t^{\prime }}\partial _{x}u\left( x,t^{\prime }\right)
\right) \,\partial _{t^{\prime }}u\left( x,t^{\prime }\right) \right]
_{x\rightarrow -\infty }^{x\rightarrow \infty }-\int_{\mathbb{R}}\left( \dot{%
\sigma}_{sr}\left( t^{\prime }\right) \ast _{t^{\prime }}\partial
_{x}u\left( x,t^{\prime }\right) \right) \,\partial _{t^{\prime }x}u\left(
x,t^{\prime }\right) \,\mathrm{d}x\right) \,\mathrm{d}t^{\prime } \\
&&\qquad =\int_{0}^{t}\!\!\int_{\mathbb{R}}\left( \left( -\dot{\sigma}%
_{sr}\left( t^{\prime }\right) \right) \ast _{t^{\prime }}\partial
_{x}u\left( x,t^{\prime }\right) \right) \,\partial _{t^{\prime }}\left(
\partial _{x}u\left( x,t^{\prime }\right) \right) \,\mathrm{d}x\,\mathrm{d}%
t^{\prime } \\
&&\qquad =\int_{\mathbb{R}}\left( \left[ \left( \left( -\dot{\sigma}%
_{sr}\left( t^{\prime }\right) \right) \ast _{t^{\prime }}\partial
_{x}u\left( x,t^{\prime }\right) \right) \,\partial _{x}u\left( x,t^{\prime
}\right) \right] _{t^{\prime }=0}^{t^{\prime }=t}-\int_{0}^{t}\partial
_{t^{\prime }}\left( \left( -\dot{\sigma}_{sr}\left( t^{\prime }\right)
\right) \ast _{t^{\prime }}\partial _{x}u\left( x,t^{\prime }\right) \right)
\,\partial _{x}u\left( x,t^{\prime }\right) \,\mathrm{d}t^{\prime }\right) \,%
\mathrm{d}x \\
&&\qquad =\int_{\mathbb{R}}\left( \left( -\dot{\sigma}_{sr}\left( t\right)
\right) \ast _{t}\partial _{x}u\left( x,t\right) \right) \,\partial
_{x}u\left( x,t\right) \,\mathrm{d}x-\int_{0}^{t}\!\!\int_{\mathbb{R}%
}\partial _{t^{\prime }}\left( \left( -\dot{\sigma}_{sr}\left( t^{\prime
}\right) \right) \ast _{t^{\prime }}\partial _{x}u\left( x,t^{\prime
}\right) \right) \,\partial _{x}u\left( x,t^{\prime }\right) \,\mathrm{d}x\,%
\mathrm{d}t^{\prime }
\end{eqnarray*}%
after the partial integration with respect to spatial coordinate and time,
using previously derived boundary condition $\lim_{x\rightarrow \pm \infty
}\partial _{x}u\left( x,t\right) =0,$ so that (\ref{ee-sigma-g-skoro}) reads%
\begin{eqnarray}
&&\frac{1}{2}\rho \left\Vert \partial _{t}u\left( \cdot ,t\right)
\right\Vert _{L^{2}\left( 
\mathbb{R}
\right) }^{2}+\frac{1}{2}\sigma _{sr}^{\left( g\right) }\left\Vert \partial
_{x}u\left( \cdot ,t\right) \right\Vert _{L^{2}\left( 
\mathbb{R}
\right) }^{2}+\int_{0}^{t}\!\!\int_{\mathbb{R}}\partial _{t^{\prime }}\left(
\left( -\dot{\sigma}_{sr}\left( t^{\prime }\right) \right) \ast _{t^{\prime
}}\partial _{x}u\left( x,t^{\prime }\right) \right) \,\partial _{x}u\left(
x,t^{\prime }\right) \,\mathrm{d}x\,\mathrm{d}t^{\prime }  \notag \\
&&\qquad \qquad \qquad =\frac{1}{2}\rho \left\Vert v_{0}\left( \cdot \right)
\right\Vert _{L^{2}\left( 
\mathbb{R}
\right) }^{2}+\int_{\mathbb{R}}\left( \left( -\dot{\sigma}_{sr}\left(
t\right) \right) \ast _{t}\partial _{x}u\left( x,t\right) \right) \,\partial
_{x}u\left( x,t\right) \,\mathrm{d}x.  \label{ee-sigma-g-skoro-1}
\end{eqnarray}

Using Lemma 1.7.2 in \cite{Siskova-phd}, see also \cite[Eq. (9)]{Zacher},
stating that 
\begin{equation}
\int_{0}^{t}\!\!\int_{\mathbb{R}}\partial _{t^{\prime }}\left( k\left(
t^{\prime }\right) \ast _{t^{\prime }}u\left( x,t^{\prime }\right) \right)
\,u\left( x,t^{\prime }\right) \,\mathrm{d}x\,\mathrm{d}t^{\prime }\geq 
\frac{1}{2}k\left( t\right) \ast _{t}\left\Vert u\left( \cdot ,t\right)
\right\Vert _{L^{2}\left( 
\mathbb{R}
\right) }^{2}+\int_{0}^{t}k\left( t^{\prime }\right) \left\Vert u\left(
\cdot ,t^{\prime }\right) \right\Vert _{L^{2}\left( 
\mathbb{R}
\right) }^{2}\,\mathrm{d}t^{\prime },  \label{lema}
\end{equation}%
provided that $k$ is a positive decreasing function for $t>0,$ the third
term on the left-hand-side of (\ref{ee-sigma-g-skoro-1}) is estimated by%
\begin{eqnarray*}
&&\int_{0}^{t}\!\!\int_{\mathbb{R}}\partial _{t^{\prime }}\left( \left( -%
\dot{\sigma}_{sr}\left( t^{\prime }\right) \right) \ast _{t^{\prime
}}\partial _{x}u\left( x,t^{\prime }\right) \right) \,\partial _{x}u\left(
x,t^{\prime }\right) \,\mathrm{d}x\,\mathrm{d}t^{\prime } \\
&&\qquad \qquad \geq \frac{1}{2}\left( -\dot{\sigma}_{sr}\left( t\right)
\right) \ast _{t}\left\Vert \partial _{x}u\left( \cdot ,t\right) \right\Vert
_{L^{2}\left( 
\mathbb{R}
\right) }^{2}+\frac{1}{2}\int_{0}^{t}\left( -\dot{\sigma}_{sr}\left(
t^{\prime }\right) \right) \left\Vert \partial _{x}u\left( \cdot ,t^{\prime
}\right) \right\Vert _{L^{2}\left( 
\mathbb{R}
\right) }^{2}\,\mathrm{d}t^{\prime },
\end{eqnarray*}%
since $-\dot{\sigma}_{sr}$ is completely monotonic and thus positive
decreasing function for $t>0,$ while the second term on the right-hand-side
of (\ref{ee-sigma-g-skoro-1}) is estimated by%
\begin{eqnarray*}
&&\int_{\mathbb{R}}\left( \left( -\dot{\sigma}_{sr}\left( t\right) \right)
\ast _{t}\partial _{x}u\left( x,t\right) \right) \,\partial _{x}u\left(
x,t\right) \,\mathrm{d}x \\
&&\qquad \qquad =\int_{0}^{t}\left( -\dot{\sigma}_{sr}\left( t-t^{\prime
}\right) \right) \int_{\mathbb{R}}\partial _{x}u\left( x,t^{\prime }\right)
\,\partial _{x}u\left( x,t\right) \,\mathrm{d}x\,\mathrm{d}t^{\prime } \\
&&\qquad \qquad \leq \int_{0}^{t}\left( -\dot{\sigma}_{sr}\left( t-t^{\prime
}\right) \right) \int_{\mathbb{R}}\left( \frac{\left( \partial _{x}u\left(
x,t^{\prime }\right) \right) ^{2}}{2}+\frac{\left( \partial _{x}u\left(
x,t\right) \right) ^{2}}{2}\right) \,\mathrm{d}x\,\mathrm{d}t^{\prime } \\
&&\qquad \qquad \leq \frac{1}{2}\left( -\dot{\sigma}_{sr}\left( t\right)
\right) \ast _{t}\left\Vert \partial _{x}u\left( \cdot ,t\right) \right\Vert
_{L^{2}\left( 
\mathbb{R}
\right) }^{2}+\frac{1}{2}\left( \sigma _{sr}^{\left( g\right) }-\sigma
_{sr}\left( t\right) \right) \left\Vert \partial _{x}u\left( \cdot ,t\right)
\right\Vert _{L^{2}\left( 
\mathbb{R}
\right) }^{2},
\end{eqnarray*}%
transforming (\ref{ee-sigma-g-skoro-1}) into%
\begin{equation}
\frac{1}{2}\rho \left\Vert \partial _{t}u\left( \cdot ,t\right) \right\Vert
_{L^{2}\left( 
\mathbb{R}
\right) }^{2}+\frac{1}{2}\sigma _{sr}\left( t\right) \left\Vert \partial
_{x}u\left( \cdot ,t\right) \right\Vert _{L^{2}\left( 
\mathbb{R}
\right) }^{2}+\frac{1}{2}\int_{0}^{t}\left( -\dot{\sigma}_{sr}\left(
t^{\prime }\right) \right) \left\Vert \partial _{x}u\left( \cdot ,t^{\prime
}\right) \right\Vert _{L^{2}\left( 
\mathbb{R}
\right) }^{2}\,\mathrm{d}t^{\prime }\leq \frac{1}{2}\rho \left\Vert
v_{0}\left( \cdot \right) \right\Vert _{L^{2}\left( 
\mathbb{R}
\right) }^{2}.  \label{ee-sigma-g}
\end{equation}

The energy estimate (\ref{ee-sigma-g}) clearly indicates the dissipativity
of fractional wave equation (\ref{FWE-sigma-g}), since the kinetic energy at
any time-instant $t>0$ is less than the kinetic energy at initial
time-instant $t=0,$ due to the positive terms on the left-hand-side of
energy estimate (\ref{ee-sigma-g}).

\subsection{Materials having infinite glass modulus}

The energy estimate for fractional wave equation expressed in terms of
relaxation modulus (\ref{FWE-sigma}) correspond to materials that have
infinite glass modulus and thus infinite wave speed as well, i.e., materials
described by Case II (\ref{Case 2}) and Case IV (\ref{Case 4}) of the linear
constitutive model (\ref{gen-lin}), as well as materials described by the
fractional Burgers models I - V (\ref{Model 1}), (\ref{Model 2}), (\ref%
{Model 3}), (\ref{Model 4}), (\ref{Model 5}).

Namely, by multiplying the fractional wave equation (\ref{FWE-sigma}) by $%
\partial _{t}u$ and by subsequent integration with respect to the spatial
coordinate along the whole domain $%
\mathbb{R}
$ and with respect to time over interval $\left[ 0,t\right] ,$ one has%
\begin{equation}
\frac{1}{2}\rho \left\Vert \partial _{t}u\left( \cdot ,t\right) \right\Vert
_{L^{2}\left( 
\mathbb{R}
\right) }^{2}=\frac{1}{2}\rho \left\Vert v_{0}\left( \cdot \right)
\right\Vert _{L^{2}\left( 
\mathbb{R}
\right) }^{2}+\int_{0}^{t}\!\!\int_{\mathbb{R}}\left( \sigma _{sr}\left(
t^{\prime }\right) \ast _{t^{\prime }}\partial _{t^{\prime }xx}u\left(
x,t^{\prime }\right) \right) \,\partial _{t^{\prime }}u\left( x,t^{\prime
}\right) \,\mathrm{d}x\,\mathrm{d}t^{\prime },  \label{ee-sigma-skoro}
\end{equation}%
where the change of kinetic energy is obtained according to (\ref{kin-en}).
The second term on the right-hand-side of (\ref{ee-sigma-skoro}) transforms
into%
\begin{eqnarray*}
&&\int_{0}^{t}\!\!\int_{\mathbb{R}}\left( \sigma _{sr}\left( t^{\prime
}\right) \ast _{t^{\prime }}\partial _{t^{\prime }xx}u\left( x,t^{\prime
}\right) \right) \,\partial _{t^{\prime }}u\left( x,t^{\prime }\right) \,%
\mathrm{d}x\,\mathrm{d}t^{\prime } \\
&&\qquad =\int_{0}^{t}\!\!\int_{\mathbb{R}}\partial _{x}\left( \sigma
_{sr}\left( t^{\prime }\right) \ast _{t^{\prime }}\partial _{t^{\prime
}x}u\left( x,t^{\prime }\right) \right) \,\partial _{t^{\prime }}u\left(
x,t^{\prime }\right) \,\mathrm{d}x\,\mathrm{d}t^{\prime } \\
&&\qquad =\int_{0}^{t}\left( \left[ \left( \sigma _{sr}\left( t^{\prime
}\right) \ast _{t^{\prime }}\partial _{t^{\prime }x}u\left( x,t^{\prime
}\right) \right) \,\partial _{t^{\prime }}u\left( x,t^{\prime }\right) %
\right] _{x\rightarrow -\infty }^{x\rightarrow \infty }-\int_{\mathbb{R}%
}\left( \sigma _{sr}\left( t^{\prime }\right) \ast _{t^{\prime }}\partial
_{t^{\prime }x}u\left( x,t^{\prime }\right) \right) \,\partial _{t^{\prime
}x}u\left( x,t^{\prime }\right) \,\mathrm{d}x\right) \,\mathrm{d}t^{\prime }
\\
&&\qquad =-\int_{0}^{t}\!\!\int_{\mathbb{R}}\left( \sigma _{sr}\left(
t^{\prime }\right) \ast _{t^{\prime }}\partial _{t^{\prime }x}u\left(
x,t^{\prime }\right) \right) \,\partial _{t^{\prime }x}u\left( x,t^{\prime
}\right) \,\mathrm{d}x\,\mathrm{d}t^{\prime },
\end{eqnarray*}%
after the partial integration with respect to spatial coordinate, using the
boundary condition (\ref{bc})$_{2}$ yielding $\lim_{x\rightarrow \pm \infty
}\sigma _{sr}\left( t\right) \ast _{t}\partial _{tx}u\left( x,t\right) =0,$
obtained by combining the constitutive equation (\ref{sigma-konv-1}) and
strain (\ref{strejn}), so that (\ref{ee-sigma-skoro}) reads%
\begin{equation}
\frac{1}{2}\rho \left\Vert \partial _{t}u\left( \cdot ,t\right) \right\Vert
_{L^{2}\left( 
\mathbb{R}
\right) }^{2}+\int_{0}^{t}\!\!\int_{\mathbb{R}}\left( \sigma _{sr}\left(
t^{\prime }\right) \ast _{t^{\prime }}\partial _{t^{\prime }x}u\left(
x,t^{\prime }\right) \right) \,\partial _{t^{\prime }x}u\left( x,t^{\prime
}\right) \,\mathrm{d}x\,\mathrm{d}t^{\prime }=\frac{1}{2}\rho \left\Vert
v_{0}\left( \cdot \right) \right\Vert _{L^{2}\left( 
\mathbb{R}
\right) }^{2}.  \label{ee-sigma}
\end{equation}

The energy estimate (\ref{ee-sigma}) clearly indicates the dissipativity of
fractional wave equation (\ref{FWE-sigma}), since the kinetic energy at any
time-instant $t>0$ is less than the kinetic energy at initial time-instant $%
t=0,$ due to the positivity of the second term on the right-hand-side of (%
\ref{ee-sigma}), thanks to the relaxation modulus $\sigma _{sr}$ being
completely monotonic and consequently of the positive type kernels
satisfying 
\begin{equation*}
\int_{0}^{t}\!\!\int_{0}^{t^{\prime }}\sigma _{sr}\left( t^{\prime
}-t^{\prime \prime }\right) \,\partial _{t^{\prime \prime }x}u\left(
x,t^{\prime \prime }\right) \,\partial _{t^{\prime }x}u\left( x,t^{\prime
}\right) \,\mathrm{d}t^{\prime \prime }\,\mathrm{d}t^{\prime }\geq 0,
\end{equation*}
as also used in \cite{Saedpanah}.

\subsection{Energy estimates using fractional wave equation (\protect\ref%
{FWE-epsilon-g})}

The energy estimate for fractional wave equation expressed in terms of creep
compliance (\ref{FWE-epsilon-g}) correspond to all materials described by
the power-type distributed-order model (\ref{DOCE}), as well as to all
materials described by the fractional Burgers models, since for all of these
models the glass compliance has finite value, either zero or non-zero.

Multiplying the fractional wave equation (\ref{FWE-epsilon-g}) by $\partial
_{t}u$ and by subsequent integration with respect to the spatial coordinate
along the whole domain $%
\mathbb{R}
$ and with respect to time over interval $\left[ 0,t\right] ,$ one has%
\begin{eqnarray}
&&\frac{1}{2}\rho \,\varepsilon _{cr}^{\left( g\right) }\left\Vert \partial
_{t}u\left( \cdot ,t\right) \right\Vert _{L^{2}\left( 
\mathbb{R}
\right) }^{2}+\frac{1}{2}\left\Vert \partial _{x}u\left( \cdot ,t\right)
\right\Vert _{L^{2}\left( 
\mathbb{R}
\right) }^{2}  \notag \\
&&\qquad \qquad \qquad +\rho \int_{0}^{t}\!\!\int_{\mathbb{R}}\left( \dot{%
\varepsilon}_{cr}\left( t^{\prime }\right) \ast _{t^{\prime }}\partial
_{t^{\prime }t^{\prime }}u\left( x,t^{\prime }\right) \right) \,\partial
_{t^{\prime }}u\left( x,t^{\prime }\right) \,\mathrm{d}x\,\mathrm{d}%
t^{\prime }=\frac{1}{2}\rho \,\varepsilon _{cr}^{\left( g\right) }\left\Vert
v_{0}\left( \cdot \right) \right\Vert _{L^{2}\left( 
\mathbb{R}
\right) }^{2},  \label{ee-epsilon-g-skoro}
\end{eqnarray}%
where the changes of kinetic and potential energy are obtained according to (%
\ref{kin-en}) and (\ref{pot-en}), respectively. The last term on the
left-hand-side of (\ref{ee-epsilon-g-skoro}) is calculated as%
\begin{eqnarray*}
&&\int_{0}^{t}\!\!\int_{\mathbb{R}}\left( \dot{\varepsilon}_{cr}\left(
t^{\prime }\right) \ast _{t^{\prime }}\partial _{t^{\prime }t^{\prime
}}u\left( x,t^{\prime }\right) \right) \,\partial _{t^{\prime }}u\left(
x,t^{\prime }\right) \,\mathrm{d}x\,\mathrm{d}t^{\prime } \\
&&\qquad =\int_{0}^{t}\!\!\int_{\mathbb{R}}\left( \partial _{t^{\prime
}}\left( \dot{\varepsilon}_{cr}\left( t^{\prime }\right) \ast _{t^{\prime
}}\partial _{t^{\prime }}u\left( x,t^{\prime }\right) \right) -v_{0}\left(
x\right) \dot{\varepsilon}_{cr}\left( t^{\prime }\right) \right) \,\partial
_{t^{\prime }}u\left( x,t^{\prime }\right) \,\mathrm{d}x\,\mathrm{d}%
t^{\prime } \\
&&\qquad =\int_{0}^{t}\!\!\int_{\mathbb{R}}\partial _{t^{\prime }}\left( 
\dot{\varepsilon}_{cr}\left( t^{\prime }\right) \ast _{t^{\prime }}\partial
_{t^{\prime }}u\left( x,t^{\prime }\right) \right) \,\partial _{t^{\prime
}}u\left( x,t^{\prime }\right) \,\mathrm{d}x\,\mathrm{d}t^{\prime
}-\int_{0}^{t}\!\!\int_{\mathbb{R}}v_{0}\left( x\right) \,\dot{\varepsilon}%
_{cr}\left( t^{\prime }\right) \,\partial _{t^{\prime }}u\left( x,t^{\prime
}\right) \,\mathrm{d}x\mathrm{d}t^{\prime }
\end{eqnarray*}%
using $f\left( t\right) \ast _{t}\dot{g}\left( t\right) =\frac{\mathrm{d}}{%
\mathrm{d}t}\left( f\left( t\right) \ast _{t}g\left( t\right) \right)
-f\left( t\right) g\left( 0\right) ,$ transforming (\ref{ee-epsilon-g-skoro}%
) into%
\begin{eqnarray}
&&\frac{1}{2}\rho \,\varepsilon _{cr}^{\left( g\right) }\left\Vert \partial
_{t}u\left( \cdot ,t\right) \right\Vert _{L^{2}\left( 
\mathbb{R}
\right) }^{2}+\frac{1}{2}\left\Vert \partial _{x}u\left( \cdot ,t\right)
\right\Vert _{L^{2}\left( 
\mathbb{R}
\right) }^{2}+\rho \int_{0}^{t}\!\!\int_{\mathbb{R}}\partial _{t^{\prime
}}\left( \dot{\varepsilon}_{cr}\left( t^{\prime }\right) \ast _{t^{\prime
}}\partial _{t^{\prime }}u\left( x,t^{\prime }\right) \right) \,\partial
_{t^{\prime }}u\left( x,t^{\prime }\right) \,\mathrm{d}x\,\mathrm{d}%
t^{\prime }  \notag \\
&&\qquad \qquad \qquad =\frac{1}{2}\rho \,\varepsilon _{cr}^{\left( g\right)
}\left\Vert v_{0}\left( \cdot \right) \right\Vert _{L^{2}\left( 
\mathbb{R}
\right) }^{2}+\rho \int_{0}^{t}\!\!\int_{\mathbb{R}}v_{0}\left( x\right) \,%
\dot{\varepsilon}_{cr}\left( t^{\prime }\right) \,\partial _{t^{\prime
}}u\left( x,t^{\prime }\right) \,\mathrm{d}x\,\mathrm{d}t^{\prime }.
\label{ee-epsilon-g-skoro-1}
\end{eqnarray}%
The last term on the left-hand-side of (\ref{ee-epsilon-g-skoro-1}) is
estimated as 
\begin{eqnarray*}
&&\int_{0}^{t}\!\!\int_{\mathbb{R}}\partial _{t^{\prime }}\left( \dot{%
\varepsilon}_{cr}\left( t^{\prime }\right) \ast _{t^{\prime }}\partial
_{t^{\prime }}u\left( x,t^{\prime }\right) \right) \,\partial _{t^{\prime
}}u\left( x,t^{\prime }\right) \,\mathrm{d}x\,\mathrm{d}t^{\prime } \\
&&\qquad \qquad \geq \frac{1}{2}\dot{\varepsilon}_{cr}\left( t\right) \ast
_{t}\left\Vert \partial _{t}u\left( \cdot ,t\right) \right\Vert
_{L^{2}\left( 
\mathbb{R}
\right) }^{2}+\frac{1}{2}\int_{0}^{t}\dot{\varepsilon}_{cr}\left( t^{\prime
}\right) \left\Vert \partial _{t^{\prime }}u\left( \cdot ,t^{\prime }\right)
\right\Vert _{L^{2}\left( 
\mathbb{R}
\right) }^{2}\,\mathrm{d}t^{\prime },
\end{eqnarray*}%
according to (\ref{lema}), since $\dot{\varepsilon}_{cr}$ is completely
monotonic, while the second term on the right-hand-side of (\ref%
{ee-epsilon-g-skoro-1}) is estimated by%
\begin{eqnarray*}
&&\int_{0}^{t}\!\!\int_{\mathbb{R}}v_{0}\left( x\right) \,\dot{\varepsilon}%
_{cr}\left( t^{\prime }\right) \,\partial _{t^{\prime }}u\left( x,t^{\prime
}\right) \,\mathrm{d}x\,\mathrm{d}t^{\prime } \\
&&\qquad \qquad =\int_{0}^{t}\dot{\varepsilon}_{cr}\left( t^{\prime }\right)
\int_{\mathbb{R}}v_{0}\left( x\right) \,\,\partial _{t^{\prime }}u\left(
x,t^{\prime }\right) \,\mathrm{d}x\,\mathrm{d}t^{\prime } \\
&&\qquad \qquad \leq \int_{0}^{t}\dot{\varepsilon}_{cr}\left( t^{\prime
}\right) \int_{\mathbb{R}}\left( \frac{\left( v_{0}\left( x\right) \right)
^{2}}{2}+\frac{\left( \partial _{t^{\prime }}u\left( x,t^{\prime }\right)
\right) ^{2}}{2}\right) \,\mathrm{d}x\,\mathrm{d}t^{\prime } \\
&&\qquad \qquad \leq \frac{1}{2}\left( \varepsilon _{cr}\left( t\right)
-\varepsilon _{cr}^{\left( g\right) }\right) \left\Vert v_{0}\left( \cdot
\right) \right\Vert _{L^{2}\left( 
\mathbb{R}
\right) }^{2}+\frac{1}{2}\int_{0}^{t}\dot{\varepsilon}_{cr}\left( t^{\prime
}\right) \left\Vert \partial _{t^{\prime }}u\left( \cdot ,t^{\prime }\right)
\right\Vert _{L^{2}\left( 
\mathbb{R}
\right) }^{2}\,\mathrm{d}t^{\prime },
\end{eqnarray*}%
transforming (\ref{ee-epsilon-g-skoro-1}) into%
\begin{equation*}
0\leq \frac{1}{2}\rho \,\varepsilon _{cr}^{\left( g\right) }\left\Vert
\partial _{t}u\left( \cdot ,t\right) \right\Vert _{L^{2}\left( 
\mathbb{R}
\right) }^{2}+\frac{1}{2}\rho \,\dot{\varepsilon}_{cr}\left( t\right) \ast
_{t}\left\Vert \partial _{t}u\left( \cdot ,t\right) \right\Vert
_{L^{2}\left( 
\mathbb{R}
\right) }^{2}+\frac{1}{2}\left\Vert \partial _{x}u\left( \cdot ,t\right)
\right\Vert _{L^{2}\left( 
\mathbb{R}
\right) }^{2}\leq \frac{1}{2}\rho \,\varepsilon _{cr}\left( t\right)
\left\Vert v_{0}\left( \cdot \right) \right\Vert _{L^{2}\left( 
\mathbb{R}
\right) }^{2},
\end{equation*}%
or equivalently to%
\begin{equation}
0\leq \frac{1}{2}\rho \,\frac{1}{\varepsilon _{cr}\left( t\right) }\partial
_{t}\left( \varepsilon _{cr}\left( t\right) \ast _{t}\left\Vert \partial
_{t}u\left( \cdot ,t\right) \right\Vert _{L^{2}\left( 
\mathbb{R}
\right) }^{2}\right) +\frac{1}{2\varepsilon _{cr}\left( t\right) }\left\Vert
\partial _{x}u\left( \cdot ,t\right) \right\Vert _{L^{2}\left( 
\mathbb{R}
\right) }^{2}\leq \frac{1}{2}\rho \,\left\Vert v_{0}\left( \cdot \right)
\right\Vert _{L^{2}\left( 
\mathbb{R}
\right) }^{2},  \label{ee-epsilon-g}
\end{equation}%
using $\dot{f}\left( t\right) \ast _{t}g\left( t\right) =\frac{\mathrm{d}}{%
\mathrm{d}t}\left( f\left( t\right) \ast _{t}g\left( t\right) \right)
-f\left( 0\right) g\left( t\right) .$

The energy estimate (\ref{ee-epsilon-g}) is not appropriate for showing
dissipativity of the fractional wave equation (\ref{FWE-epsilon-g}), since
one cannot identify the kinetic energy on the left-hand-side of (\ref%
{ee-epsilon-g}), although it figures on the right-hand-side of (\ref%
{ee-epsilon-g}).

\section{Energy conservation for non-local materials}

A priori energy estimates yield the conservation law for both of the
examined non-local fractional wave equations, stating that the sum of
kinetic energy and non-local potential energy does not change in time.
Non-local potential energy is proportional to the square of fractional
strain, obtained by convoluting the classical strain with the constitutive
model dependant non-locality kernel, i.e., non-local potential energy in a
particular point depends on the square of strain in all other points
weighted by the non-locality kernel.

\subsection{Materials described by the non-local Hooke law}

Eliminating stress and strain from the equation of motion (\ref{eq-motion}),
non-local Hooke law (\ref{nl-Huk}), and strain (\ref{strejn}), the non-local
Hooke-type wave equation is obtained in the form%
\begin{equation}
\rho \,\partial _{tt}u(x,t)=\frac{E}{\ell ^{1-\alpha }}\,\frac{|x|^{-\alpha }%
}{2\Gamma (1-\alpha )}\ast _{x}\partial _{xx}u(x,t),\;\;\alpha \in \left(
0,1\right) ,  \label{AS}
\end{equation}%
transforming into%
\begin{equation}
\rho \,\partial _{tt}\hat{u}(\xi ,t)=-E\frac{\sin \frac{\alpha \pi }{2}}{%
\ell ^{1-\alpha }}|\xi |^{1+\alpha }\hat{u}(\xi ,t),  \label{AS-ft}
\end{equation}%
after application of the Fourier transform with respect to the spatial
coordinate%
\begin{equation*}
\hat{f}\left( \xi \right) =\mathcal{F}\left[ f\left( x\right) \right] \left(
\xi \right) =\int_{%
\mathbb{R}
}f\left( x\right) \mathrm{e}^{-\mathrm{i}\xi x}\,\mathrm{d}x,\;\;\xi \in 
\mathbb{R}
,
\end{equation*}%
where $\mathcal{F}\left[ \frac{|x|^{-\alpha }}{2\Gamma (1-\alpha )}\right]
\left( \xi \right) =\frac{\sin \frac{\alpha \pi }{2}}{|\xi |^{1-\alpha }}$
is used along with other well-known properties of the Fourier transform.

Multiplying the non-local Hooke-type wave equation in Fourier domain (\ref%
{AS-ft}) with $\partial _{t}\hat{u}$ and by subsequent integration over the
whole domain $%
\mathbb{R}
,$ one obtains%
\begin{equation}
\partial _{t}\left( \frac{1}{2}\rho \left\Vert \partial _{t}\hat{u}(\cdot
,t)\right\Vert _{L^{2}\left( 
\mathbb{R}
\right) }^{2}+\frac{1}{2}E\frac{\sin \frac{\alpha \pi }{2}}{\ell ^{1-\alpha }%
}\left\Vert |\xi |^{\frac{1+\alpha }{2}}\hat{u}(\xi ,t)\right\Vert
_{L^{2}\left( 
\mathbb{R}
\right) }^{2}\right) =0,  \label{ZO-Htajp}
\end{equation}%
yielding the conservation law%
\begin{gather}
\partial _{t}\left( \frac{1}{2}\rho \left\Vert \partial _{t}u(\cdot
,t)\right\Vert _{L^{2}\left( 
\mathbb{R}
\right) }^{2}+\frac{1}{2}E\frac{\sin \frac{\alpha \pi }{2}}{\ell ^{1-\alpha }%
}\left\Vert \left( -\Delta \right) ^{\frac{1+\alpha }{4}}u\left( \cdot
,t\right) \right\Vert _{L^{2}\left( 
\mathbb{R}
\right) }^{2}\right) =0,\;\;\text{i.e.,}  \notag \\
\frac{1}{2}\rho \left\Vert \partial _{t}u(\cdot ,t)\right\Vert _{L^{2}\left( 
\mathbb{R}
\right) }^{2}+\frac{1}{2}E\frac{\sin \frac{\alpha \pi }{2}}{\ell ^{1-\alpha }%
}\left\Vert \left( -\Delta \right) ^{\frac{1+\alpha }{4}}u\left( \cdot
,t\right) \right\Vert _{L^{2}\left( 
\mathbb{R}
\right) }^{2}=\mathrm{const.},  \label{ZO-AS}
\end{gather}%
by the Parseval identity $\left\Vert f\right\Vert _{L^{2}\left( 
\mathbb{R}
\right) }^{2}=\left\Vert \hat{f}\right\Vert _{L^{2}\left( 
\mathbb{R}
\right) }^{2},$ as well as by the Fourier transform of fractional Laplacian
(in one dimension) $\mathcal{F}\left[ \left( -\Delta \right) ^{s}f\left(
x\right) \right] \left( \xi \right) =|\xi |^{2s}\hat{f}\left( \xi \right) ,$
with $s\in \left( 0,1\right) ,$ since $\frac{1+\alpha }{2}\in \left( \frac{1%
}{2},1\right) $. The fractional strain, being proportional to $\left(
-\Delta \right) ^{\frac{1+\alpha }{4}}u$ in (\ref{ZO-AS}), has a lower
differentiation order than the classical strain $\partial _{x}u,$ since $%
\frac{1+\alpha }{4}\in \left( \frac{1}{4},\frac{1}{2}\right) $.

However, the conservation law (\ref{ZO-AS}) may also take another form%
\begin{gather}
\partial _{t}\left( \frac{1}{2}\rho \left\Vert \partial _{t}u(\cdot
,t)\right\Vert _{L^{2}\left( 
\mathbb{R}
\right) }^{2}+\frac{1}{2}E\frac{\sin \frac{\alpha \pi }{2}}{2\ell ^{1-\alpha
}\Gamma \left( 1-\frac{1+\alpha }{2}\right) \cos \frac{\left( 1+\alpha
\right) \pi }{4}}\left\Vert \frac{\mathrm{sgn\,}x}{|x|^{\frac{1+\alpha }{2}}}%
\ast _{x}\partial _{x}u\left( x,t\right) \right\Vert _{L^{2}\left( 
\mathbb{R}
\right) }^{2}\right) =0,\;\;\text{i.e.,}  \notag \\
\frac{1}{2}\rho \left\Vert \partial _{t}u(\cdot ,t)\right\Vert _{L^{2}\left( 
\mathbb{R}
\right) }^{2}+\frac{1}{2}E\frac{c_{\alpha }}{\ell ^{1-\alpha }}\left\Vert 
\frac{\mathrm{sgn\,}x}{|x|^{\frac{1+\alpha }{2}}}\ast _{x}\partial
_{x}u\left( x,t\right) \right\Vert _{L^{2}\left( 
\mathbb{R}
\right) }^{2}=\mathrm{const}.  \label{ZO-AS-1}
\end{gather}%
where $c_{\alpha }=\frac{\sin \frac{\alpha \pi }{2}}{2\Gamma \left( 1-\frac{%
1+\alpha }{2}\right) \cos \frac{\left( 1+\alpha \right) \pi }{4}}$ is a
positive constant, if the term $|\xi |^{\frac{1+\alpha }{2}}\hat{u}(\xi ,t)$
in (\ref{ZO-Htajp}) is rewritten as%
\begin{equation*}
|\xi |^{\frac{1+\alpha }{2}}\hat{u}(\xi ,t)=-\mathrm{i}\frac{\mathrm{sgn\,}%
\xi }{|\xi |^{1-\frac{1+\alpha }{2}}}\left( \mathrm{i}\xi \,\hat{u}(\xi
,t)\right) =\frac{1}{2\Gamma \left( 1-\frac{1+\alpha }{2}\right) \cos \frac{%
\left( 1+\alpha \right) \pi }{4}}\mathcal{F}\left[ \frac{\mathrm{sgn\,}x}{%
|x|^{\frac{1+\alpha }{2}}}\right] \left( \xi \right) \,\mathcal{F}\left[
\partial _{x}u\left( x,t\right) \right] \left( \xi \right) ,
\end{equation*}%
where the Fourier transform $\mathcal{F}\left[ \frac{\mathrm{sgn\,}x}{%
|x|^{\beta }}\right] \left( \xi \right) =-2\mathrm{i}\Gamma \left( 1-\beta
\right) \cos \frac{\beta \pi }{2}\frac{\mathrm{sgn\,}\xi }{|\xi |^{1-\beta }}%
,$ with $\beta \in \left( 0,1\right) ,$ is used.

The energy estimates (\ref{ZO-AS}) and (\ref{ZO-AS-1}) clearly indicate the
energy conservation property of the non-local Hooke-type wave equation (\ref%
{AS}), if the potential energy is reinterpreted to be proportional to the
square of fractional strain, expressed either in terms of fractional
Laplacian, or in terms of classical strain convoluted by the non-locality
kernel of power type.

\subsection{Materials described by the fractional Eringen model}

Fractional Eringen wave equation%
\begin{eqnarray}
&&\rho \,\partial _{tt}u(x,t)=E\,H_{\alpha }\left( x\right) \ast
_{x}\partial _{xx}u\left( x,t\right) ,\;\;\alpha \in \left( 1,3\right) ,\;\;%
\text{with}  \label{EringenWE} \\
&&H_{\alpha }\left( x\right) =\frac{1}{\pi }\int_{0}^{\infty }\frac{\cos
\left( \xi x\right) }{1+\left( \ell \xi \right) ^{\alpha }\left\vert \cos 
\frac{\alpha \pi }{2}\right\vert }\mathrm{d}\xi =\mathcal{F}^{-1}\left[ 
\frac{1}{1+\left( \ell \left\vert \xi \right\vert \right) ^{\alpha
}\left\vert \cos \frac{\alpha \pi }{2}\right\vert }\right] \left( x\right) ,
\notag
\end{eqnarray}%
is found as the inverse Fourier transform of 
\begin{equation}
\rho \,\partial _{tt}\hat{u}(\xi ,t)=-E\frac{\xi ^{2}}{1+\left( \ell
\left\vert \xi \right\vert \right) ^{\alpha }\left\vert \cos \frac{\alpha
\pi }{2}\right\vert }\hat{u}(\xi ,t),  \label{Eringen-ft}
\end{equation}%
obtained by eliminating $\hat{\sigma}$ and $\hat{\varepsilon}$ from the
system of equations in the Fourier domain 
\begin{gather*}
\mathrm{i}\xi \,\hat{\sigma}(\xi ,t)=\rho \,\partial _{tt}\hat{u}(\xi
,t),\;\;\hat{\varepsilon}\left( \xi ,t\right) =\mathrm{i}\xi \,\hat{u}(\xi
,t), \\
\left( 1+\left( \ell \left\vert \xi \right\vert \right) ^{\alpha }\left\vert
\cos \frac{\alpha \pi }{2}\right\vert \right) \hat{\sigma}(\xi ,t)=E\,\hat{%
\varepsilon}\left( \xi ,t\right) ,
\end{gather*}%
respectively consisting of the Fourier transforms of equation of motion (\ref%
{eq-motion}), strain (\ref{strejn}), and fractional Eringen model (\ref%
{frac-Eringen}), where the Fourier transform of both (\ref{Dx-1}) and (\ref%
{Dx-2}), yielding $\mathcal{F}\left[ \mathrm{D}_{x}^{\alpha }f\left(
x\right) \right] \left( \xi \right) =-\left\vert \xi \right\vert ^{\alpha
}\left\vert \cos \frac{\alpha \pi }{2}\right\vert \hat{f}\left( \xi \right)
, $ is used.

Multiplying the fractional Eringen wave equation in Fourier domain (\ref%
{Eringen-ft}) with $\partial _{t}\hat{u}$ and by subsequent integration over
the whole domain $%
\mathbb{R}
,$ one obtains%
\begin{eqnarray}
&&\partial _{t}\left( \frac{1}{2}\rho \left\Vert \partial _{t}\hat{u}(\cdot
,t)\right\Vert _{L^{2}\left( 
\mathbb{R}
\right) }^{2}+\frac{1}{2}E\left\Vert \hat{h}_{\alpha }\left( \xi \right)
\left( \mathrm{i}\xi \,\hat{u}(\xi ,t)\right) \right\Vert _{L^{2}\left( 
\mathbb{R}
\right) }^{2}\right) 
\begin{tabular}{l}
=%
\end{tabular}%
0,\;\;\text{with}  \label{Eringen-CL-ft} \\
&&\hat{h}_{\alpha }\left( \xi \right) =-\mathrm{i}\frac{\mathrm{sgn\,}\xi }{%
\sqrt{1+\left( \ell \left\vert \xi \right\vert \right) ^{\alpha }\left\vert
\cos \frac{\alpha \pi }{2}\right\vert }},  \notag
\end{eqnarray}%
so that the conservation law 
\begin{gather}
\partial _{t}\left( \frac{1}{2}\rho \left\Vert \partial _{t}u(\cdot
,t)\right\Vert _{L^{2}\left( 
\mathbb{R}
\right) }^{2}+\frac{1}{2}E\left\Vert h_{\alpha }\left( x\right) \ast
_{x}\partial _{x}u\left( x,t\right) \right\Vert _{L^{2}\left( 
\mathbb{R}
\right) }^{2}\right) =0,\;\;\text{i.e.,}  \notag \\
\frac{1}{2}\rho \left\Vert \partial _{t}u(\cdot ,t)\right\Vert _{L^{2}\left( 
\mathbb{R}
\right) }^{2}+\frac{1}{2}E\left\Vert h_{\alpha }\left( x\right) \ast
_{x}\partial _{x}u\left( x,t\right) \right\Vert _{L^{2}\left( 
\mathbb{R}
\right) }^{2}=\mathrm{const}.  \label{ZO-Eringen}
\end{gather}%
follows from (\ref{Eringen-CL-ft}) by the Parseval identity and inverse
Fourier transform of $\hat{h}_{\alpha },$ given by%
\begin{equation*}
h_{\alpha }\left( x\right) =\frac{1}{\pi }\int_{0}^{\infty }\frac{\sin
\left( \xi x\right) }{\sqrt{1+\left( \ell \xi \right) ^{\alpha }\left\vert
\cos \frac{\alpha \pi }{2}\right\vert }}\mathrm{d}\xi .
\end{equation*}

The energy estimate (\ref{ZO-Eringen}) clearly indicates the energy
conservation property of the fractional Eringen wave equation (\ref%
{EringenWE}), if the potential energy is again reinterpreted to be
proportional to the square of fractional strain, expressed in terms of
classical strain convoluted by the non-locality kernel $h_{\alpha }.$

\section{Conclusion}

Energy dissipation and conservation properties of fractional wave equations,
respectively corresponding to hereditary and non-local materials, are
considered by employing the method of a priori energy estimates. More
precisely, in the case of hereditary fractional wave equations it is
obtained that the kinetic energy at arbitrary time-instant is less than the
initial kinetic energy, while in the case of non-local fractional wave
equations it is obtained that the sum of kinetic energy and non-local
potential energy does not change in time, with the non-local potential
energy being proportional to the square of fractional strain, obtained by
convoluting the classical strain with the constitutive model dependant
non-locality kernel.

Hereditary fractional models of viscoelastic material having differentiation
orders below the first order are represented by the distributed-order
viscoelastic model (\ref{const-eq}), more precisely by the linear fractional
model (\ref{gen-lin}) and power-type distributed-order model (\ref{DOCE}),
while thermodynamically consistent fractional Burgers models (\ref{UCE-1-5})
and (\ref{UCE-6-8}) represent constitutive models having differentiation
orders up to the second order. In order to formulate the hereditary wave
equation, in addition to the equation of motion (\ref{eq-motion}) and strain
(\ref{strejn}), hereditary constitutive model expressed in terms of material
response in stress relaxation and creep test is used, leading to six
equivalent forms of the hereditary wave equation, three of them expressed in
terms of relaxation modulus and the other three expressed in terms of creep
compliance. It is found that the hereditary wave equation expressed in terms
of relaxation modulus, either as (\ref{FWE-sigma-g}) for materials having
finite glass modulus and thus finite wave speed as well, or as (\ref%
{FWE-sigma}) for materials having infinite glass modulus and thus infinite
wave speed as well, leads to the physically meaningful energy estimates
either (\ref{ee-sigma-g}) or (\ref{ee-sigma}) corresponding to energy
dissipation. Therefore, the form of energy estimate depends on the material
properties at the initial time-instant defining the wave propagation speed,
rather than the material properties for large time differing the solid- and
fluid-like materials. The energy estimate (\ref{ee-epsilon-g}), implied by
the hereditary wave equation expressed in terms of creep compliance (\ref%
{FWE-epsilon-g}), did not prove to have physical meaning.

Monotonicity property of the relaxation modulus, being completely monotonic
function and creep compliance, being Bernstein function, is the key point in
proving dissipativity properties of the hereditary fractional wave
equations. It is shown that the requirement for relaxation modulus to be
completely monotonic, i.e., creep compliance to be Bernstein function, is
equivalent with the thermodynamical conditions for linear fractional model (%
\ref{gen-lin}) and power-type distributed-order model (\ref{DOCE}), while in
the case of the fractional Burgers models these monotonicity requirements
are more restrictive than the thermodynamical requirements, as found in \cite%
{OZ-2}.

Non-local Hooke and Eringen fractional wave equations, given by (\ref{AS})
and (\ref{EringenWE}), are respectively obtained by coupling non-local
constitutive models of Hooke- and Eringen-type, (\ref{nl-Huk}) and (\ref%
{frac-Eringen}), with the equation of motion (\ref{eq-motion}) and strain (%
\ref{strejn}). A priori energy estimates (\ref{ZO-AS}) and (\ref{ZO-AS-1})
for non-local Hooke and energy estimate (\ref{ZO-Eringen}) for the
fractional Eringen wave equation imply the energy conservation, with the
reinterpreted notion of the potential energy, being in a particular point
dependant on the square of strain in all other points weighted by the model
dependent non-locality kernel. In particular, in the energy estimates (\ref%
{ZO-AS}) the non-local potential energy is proportional to the fractional
strain, represented by the action of fractional Laplacian on the
displacement.

\appendix

\section{Thermodynamically consistent linear fractional models \label{LFMS}}

Linear fractional model (\ref{gen-lin}), containing fractional
differentiation orders below the first order, reduces to the following four
thermodynamically consistent model classes, which are listed below, along
with corresponding thermodynamical constraints and explicit forms of
function $K,$ given by (\ref{K}).

\noindent \textbf{Case I}: Models having the same number and orders of
fractional derivatives of stress and strain:%
\begin{gather}
\sum_{i=1}^{n}a_{i}\,{}_{0}\mathrm{D}_{t}^{\alpha _{i}}\sigma \left(
t\right) =\sum_{i=1}^{n}b_{i}\,{}_{0}\mathrm{D}_{t}^{\alpha _{i}}\varepsilon
\left( t\right) ,  \label{Case 1} \\
0\leq \alpha _{1}<\ldots <\alpha _{n}<1\;\;\text{and}\;\;\frac{a_{1}}{b_{1}}%
\geq \frac{a_{2}}{b_{2}}\geq \ldots \geq \frac{a_{n}}{b_{n}}\geq 0,  \notag
\\
K\left( \rho \right)
=-\sum_{i=1}^{n}\sum_{j=i+1}^{n}(a_{i}b_{j}-a_{j}b_{i})\rho ^{\alpha
_{i}+\alpha _{j}}\sin \frac{(\alpha _{i}-\alpha _{j})\pi }{2};  \notag
\end{gather}

\noindent \textbf{Case II}: Models having some extra derivatives of strain
in addition to the same number and orders of fractional derivatives of
stress and strain:%
\begin{gather}
\sum_{i=1}^{n}a_{i}\,{}_{0}\mathrm{D}_{t}^{\alpha _{i}}\sigma \left(
t\right) =\sum_{i=1}^{n}b_{i}\,{}_{0}\mathrm{D}_{t}^{\alpha _{i}}\varepsilon
\left( t\right) +\sum_{i=n+1}^{m}b_{i}\,{}_{0}\mathrm{D}_{t}^{\beta
_{i}}\varepsilon \left( t\right) ,  \label{Case 2} \\
0\leq \alpha _{1}<\ldots <\alpha _{n}<\beta _{n+1}<\ldots <\beta _{m}<1\;\;%
\text{and}\;\;\frac{a_{1}}{b_{1}}\geq \frac{a_{2}}{b_{2}}\geq \ldots \geq 
\frac{a_{n}}{b_{n}}\geq 0,  \notag \\
K\left( \rho \right)
=-\sum_{i=1}^{n}\sum_{j=i+1}^{n}(a_{i}b_{j}-a_{j}b_{i})\rho ^{\alpha
_{i}+\alpha _{j}}\sin \frac{(\alpha _{i}-\alpha _{j})\pi }{2}%
-\sum_{i=1}^{n}\sum_{j=n+1}^{m}a_{i}b_{j}\,\rho ^{\alpha _{i}+\beta
_{j}}\sin \frac{(\alpha _{i}-\beta _{j})\pi }{2};  \notag
\end{gather}

\noindent \textbf{Case III}: Models having some extra derivatives of stress
in addition to the same number and orders of fractional derivatives of
stress and strain:%
\begin{gather}
\sum_{i=1}^{n-m}a_{i}\,{}_{0}\mathrm{D}_{t}^{\alpha _{i}}\sigma \left(
t\right) +\sum_{i=n-m+1}^{n}a_{i}\,{}_{0}\mathrm{D}_{t}^{\alpha _{i}}\sigma
\left( t\right) =\sum_{j=1}^{m}b_{j}\,{}_{0}\mathrm{D}_{t}^{\alpha
_{n-m+j}}\varepsilon \left( t\right) ,  \label{Case 3} \\
0\leq \alpha _{1}<\ldots <\alpha _{m}<\alpha _{m+1}<\ldots <\alpha _{n}<1\;\;%
\text{and}\;\;\frac{a_{n-m+1}}{b_{1}}\geq \frac{a_{n-m+2}}{b_{2}}\geq \ldots
\geq \frac{a_{n}}{b_{m}}\geq 0,  \notag \\
K\left( \rho \right) =-\sum_{i=1}^{n-m}\sum_{j=1}^{m}a_{i}b_{j}\,\rho
^{\alpha _{i}+\alpha _{n-m+j}}\sin \frac{(\alpha _{i}-\alpha _{n-m+j})\pi }{2%
}  \notag \\
-\frac{1}{2}\sum_{j=1}^{m}\sum_{i=n-m+j+1}^{n}\left(
a_{i}b_{j}-a_{n-m+j}b_{i-(n-m)}\right) \rho ^{\alpha _{i}+\alpha
_{n-m+j}}\sin \frac{(\alpha _{n-m+j}-\alpha _{i})\pi }{2};  \notag
\end{gather}%
\noindent \textbf{Case IV}: Models having all orders of fractional
derivatives of stress and strain different:%
\begin{gather}
\sum_{i=1}^{n}a_{i}\,{}_{0}\mathrm{D}_{t}^{\alpha _{i}}\sigma \left(
t\right) =\sum_{j=1}^{m}b_{j}\,{}_{0}\mathrm{D}_{t}^{\beta _{j}}\varepsilon
\left( t\right) ,  \label{Case 4} \\
0\leq \alpha _{1}<\ldots <\alpha _{n}<\beta _{1}<\ldots <\beta _{m}<1,\;\;%
\text{with}\;\;\alpha _{i}\not=\beta _{j},\;\;\text{for}\;\;i\not=j,  \notag
\\
K\left( \rho \right) =-\sum_{i=1}^{n}\sum_{j=1}^{m}a_{i}b_{j}\,\rho ^{\alpha
_{i}+\beta _{j}}\sin \frac{\left( \alpha _{i}-\beta _{j}\right) \pi }{2}. 
\notag
\end{gather}

\section{Fractional Burgers models \label{FBMS}}

Thermodynamically consistent fractional Burgers models are listed below,
along with corresponding thermodynamical constraints, as well as with the
constraints on monotonicity of relaxation modulus and creep compliance,
narrowing down the thermodynamical requirements and guaranteeing that
relaxation modulus is completely monotonic, while creep compliance is
Bernstein function.

\noindent \textbf{Model I}: 
\begin{gather}
\left( 1+a_{1}\,{}_{0}\mathrm{D}_{t}^{\alpha }+a_{2}\,{}_{0}\mathrm{D}%
_{t}^{\beta }+a_{3}\,{}_{0}\mathrm{D}_{t}^{\gamma }\right) \sigma \left(
t\right) =\left( b_{1}\,{}_{0}\mathrm{D}_{t}^{\mu }+b_{2}\,{}_{0}\mathrm{D}%
_{t}^{\mu +\eta }\right) \varepsilon \left( t\right) ,  \label{Model 1} \\
0\leq \alpha \leq \beta \leq \gamma \leq \mu \leq 1,\;\;1\leq \mu +\eta \leq
1+\alpha ,\;\;\frac{b_{2}}{b_{1}}\leq a_{i}\frac{\cos \frac{\left( \mu -\eta
\right) \pi }{2}}{\left\vert \cos \frac{\left( \mu +\eta \right) \pi }{2}%
\right\vert },  \notag  \label{Model 1 - tdr} \\
\frac{b_{2}}{b_{1}}\leq a_{i}\frac{\sin \frac{\left( \mu -\eta \right) \pi }{%
2}}{\sin \frac{\left( \mu +\eta \right) \pi }{2}}\frac{\cos \frac{\left( \mu
-\eta \right) \pi }{2}}{\left\vert \cos \frac{\left( \mu +\eta \right) \pi }{%
2}\right\vert },  \notag
\end{gather}%
with $\left( \eta ,i\right) \in \left\{ \left( \alpha ,1\right) ,\left(
\beta ,2\right) ,\left( \gamma ,3\right) \right\} ;$

\noindent \textbf{Model II}:%
\begin{gather}
\left( 1+a_{1}\,{}_{0}\mathrm{D}_{t}^{\alpha }+a_{2}\,{}_{0}\mathrm{D}%
_{t}^{\beta }+a_{3}\,{}_{0}\mathrm{D}_{t}^{2\alpha }\right) \sigma \left(
t\right) =\left( b_{1}\,{}_{0}\mathrm{D}_{t}^{\mu }+b_{2}\,{}_{0}\mathrm{D}%
_{t}^{\mu +\alpha }\right) \varepsilon \left( t\right) ,  \label{Model 2} \\
\frac{1}{2}\leq \alpha \leq \beta \leq \mu \leq 1,\;\;\frac{a_{3}}{a_{1}}%
\frac{\left\vert \sin \frac{\left( \mu -2\alpha \right) \pi }{2}\right\vert 
}{\sin \frac{\mu \pi }{2}}\leq \frac{b_{2}}{b_{1}}\leq a_{1}\frac{\cos \frac{%
\left( \mu -\alpha \right) \pi }{2}}{\left\vert \cos \frac{\left( \mu
+\alpha \right) \pi }{2}\right\vert },  \notag  \label{Model 2 - tdr} \\
\frac{a_{3}}{a_{1}}\frac{\left\vert \sin \frac{\left( \mu -2\alpha \right)
\pi }{2}\right\vert }{\sin \frac{\mu \pi }{2}}\frac{\cos \frac{\left( \mu
-2\alpha \right) \pi }{2}}{\cos \frac{\mu \pi }{2}}\leq \frac{b_{2}}{b_{1}}%
\leq a_{1}\frac{\sin \frac{\left( \mu -\alpha \right) \pi }{2}}{\sin \frac{%
\left( \mu +\alpha \right) \pi }{2}}\frac{\cos \frac{\left( \mu -\alpha
\right) \pi }{2}}{\left\vert \cos \frac{\left( \mu +\alpha \right) \pi }{2}%
\right\vert };  \notag
\end{gather}

\noindent \textbf{Model III}:%
\begin{gather}
\left( 1+a_{1}\,{}_{0}\mathrm{D}_{t}^{\alpha }+a_{2}\,{}_{0}\mathrm{D}%
_{t}^{\beta }+a_{3}\,{}_{0}\mathrm{D}_{t}^{\alpha +\beta }\right) \sigma
\left( t\right) =\left( b_{1}\,{}_{0}\mathrm{D}_{t}^{\mu }+b_{2}\,{}_{0}%
\mathrm{D}_{t}^{\mu +\alpha }\right) \varepsilon \left( t\right) ,
\label{Model 3} \\
0\leq \alpha \leq \beta \leq \mu \leq 1,\;\;\alpha +\beta \geq 1,\;\;\frac{%
a_{3}}{a_{2}}\frac{\left\vert \sin \frac{\left( \mu -\beta -\alpha \right)
\pi }{2}\right\vert }{\sin \frac{\left( \mu -\beta +\alpha \right) \pi }{2}}%
\leq \frac{b_{2}}{b_{1}}\leq a_{1}\frac{\cos \frac{\left( \mu -\alpha
\right) \pi }{2}}{\left\vert \cos \frac{\left( \mu +\alpha \right) \pi }{2}%
\right\vert },  \notag  \label{Model 3 - tdr} \\
\frac{a_{3}}{a_{2}}\frac{\left\vert \sin \frac{\left( \mu -\beta -\alpha
\right) \pi }{2}\right\vert }{\sin \frac{\left( \mu -\beta +\alpha \right)
\pi }{2}}\frac{\cos \frac{\left( \mu -\beta -\alpha \right) \pi }{2}}{\cos 
\frac{\left( \mu -\beta +\alpha \right) \pi }{2}}\leq \frac{b_{2}}{b_{1}}%
\leq a_{1}\frac{\sin \frac{\left( \mu -\alpha \right) \pi }{2}}{\sin \frac{%
\left( \mu +\alpha \right) \pi }{2}}\frac{\cos \frac{\left( \mu -\alpha
\right) \pi }{2}}{\left\vert \cos \frac{\left( \mu +\alpha \right) \pi }{2}%
\right\vert };  \notag
\end{gather}

\noindent \textbf{Model IV}:%
\begin{gather}
\left( 1+a_{1}\,{}_{0}\mathrm{D}_{t}^{\alpha }+a_{2}\,{}_{0}\mathrm{D}%
_{t}^{\beta }+a_{3}\,{}_{0}\mathrm{D}_{t}^{\alpha +\beta }\right) \sigma
\left( t\right) =\left( b_{1}\,{}_{0}\mathrm{D}_{t}^{\mu }+b_{2}\,{}_{0}%
\mathrm{D}_{t}^{\mu +\beta }\right) \varepsilon \left( t\right) ,
\label{Model 4} \\
0\leq \alpha \leq \beta \leq \mu \leq 1,\;\;1-\alpha \leq \beta \leq
1-\left( \mu -\alpha \right) ,\;\;\frac{a_{3}}{a_{1}}\frac{\left\vert \sin 
\frac{\left( \mu -\alpha -\beta \right) \pi }{2}\right\vert }{\sin \frac{%
\left( \mu -\alpha +\beta \right) \pi }{2}}\leq \frac{b_{2}}{b_{1}}\leq a_{2}%
\frac{\cos \frac{\left( \mu -\beta \right) \pi }{2}}{\left\vert \cos \frac{%
\left( \mu +\beta \right) \pi }{2}\right\vert },  \notag
\label{Model 4 - tdr} \\
\frac{a_{3}}{a_{1}}\frac{\left\vert \sin \frac{\left( \mu -\alpha -\beta
\right) \pi }{2}\right\vert }{\sin \frac{\left( \mu -\alpha +\beta \right)
\pi }{2}}\frac{\cos \frac{\left( \mu -\alpha -\beta \right) \pi }{2}}{\cos 
\frac{\left( \mu -\alpha +\beta \right) \pi }{2}}\leq \frac{b_{2}}{b_{1}}%
\leq a_{2}\frac{\sin \frac{\left( \mu -\beta \right) \pi }{2}}{\sin \frac{%
\left( \mu +\beta \right) \pi }{2}}\frac{\cos \frac{\left( \mu -\beta
\right) \pi }{2}}{\left\vert \cos \frac{\left( \mu +\beta \right) \pi }{2}%
\right\vert };  \notag
\end{gather}

\noindent \textbf{Model V}:%
\begin{gather}
\left( 1+a_{1}\,{}_{0}\mathrm{D}_{t}^{\alpha }+a_{2}\,{}_{0}\mathrm{D}%
_{t}^{\beta }+a_{3}\,{}_{0}\mathrm{D}_{t}^{2\beta }\right) \sigma \left(
t\right) =\left( b_{1}\,{}_{0}\mathrm{D}_{t}^{\mu }+b_{2}\,{}_{0}\mathrm{D}%
_{t}^{\mu +\beta }\right) \varepsilon \left( t\right) ,  \label{Model 5} \\
0\leq \alpha \leq \beta \leq \mu \leq 1,\;\;\frac{1}{2}\leq \beta \leq
1-\left( \mu -\alpha \right) ,\;\;\frac{a_{3}}{a_{2}}\frac{\left\vert \sin 
\frac{\left( \mu -2\beta \right) \pi }{2}\right\vert }{\sin \frac{\mu \pi }{2%
}}\leq \frac{b_{2}}{b_{1}}\leq a_{2}\frac{\cos \frac{\left( \mu -\beta
\right) \pi }{2}}{\left\vert \cos \frac{\left( \mu +\beta \right) \pi }{2}%
\right\vert },  \label{Model 5 - tdr} \\
\frac{a_{3}}{a_{2}}\frac{\left\vert \sin \frac{\left( \mu -2\beta \right)
\pi }{2}\right\vert }{\sin \frac{\mu \pi }{2}}\frac{\cos \frac{\left( \mu
-2\beta \right) \pi }{2}}{\cos \frac{\mu \pi }{2}}\leq \frac{b_{2}}{b_{1}}%
\leq a_{2}\frac{\sin \frac{\left( \mu -\beta \right) \pi }{2}}{\sin \frac{%
\left( \mu +\beta \right) \pi }{2}}\frac{\cos \frac{\left( \mu -\beta
\right) \pi }{2}}{\left\vert \cos \frac{\left( \mu +\beta \right) \pi }{2}%
\right\vert };
\end{gather}

\noindent \textbf{Model VI}:%
\begin{gather}
\left( 1+a_{1}\,{}_{0}\mathrm{D}_{t}^{\alpha }+a_{2}\,{}_{0}\mathrm{D}%
_{t}^{\beta }+a_{3}\,{}_{0}\mathrm{D}_{t}^{\alpha +\beta }\right) \sigma
\left( t\right) =\left( b_{1}\,{}_{0}\mathrm{D}_{t}^{\beta }+b_{2}\,{}_{0}%
\mathrm{D}_{t}^{\alpha +\beta }\right) \varepsilon \left( t\right) ,
\label{Model 6} \\
0\leq \alpha \leq \beta \leq 1,\;\;\alpha +\beta \geq 1,\;\;\frac{a_{3}}{%
a_{2}}\leq \frac{b_{2}}{b_{1}}\leq a_{1}\frac{\cos \frac{\left( \beta
-\alpha \right) \pi }{2}}{\left\vert \cos \frac{\left( \beta +\alpha \right)
\pi }{2}\right\vert },  \notag  \label{Model 6 - tdr} \\
\frac{a_{3}}{a_{2}}\leq \frac{b_{2}}{b_{1}}\leq a_{1}\frac{\sin \frac{\left(
\beta -\alpha \right) \pi }{2}}{\sin \frac{\left( \beta +\alpha \right) \pi 
}{2}}\frac{\cos \frac{\left( \beta -\alpha \right) \pi }{2}}{\left\vert \cos 
\frac{\left( \beta +\alpha \right) \pi }{2}\right\vert }\leq a_{1}\frac{\cos 
\frac{\left( \beta -\alpha \right) \pi }{2}}{\left\vert \cos \frac{\left(
\beta +\alpha \right) \pi }{2}\right\vert };  \notag
\end{gather}

\noindent \textbf{Model VII}:%
\begin{gather}
\left( 1+a_{1}\,{}_{0}\mathrm{D}_{t}^{\alpha }+a_{2}\,{}_{0}\mathrm{D}%
_{t}^{\beta }+a_{3}\,{}_{0}\mathrm{D}_{t}^{2\beta }\right) \sigma \left(
t\right) =\left( b_{1}\,{}_{0}\mathrm{D}_{t}^{\beta }+b_{2}\,{}_{0}\mathrm{D}%
_{t}^{2\beta }\right) \varepsilon \left( t\right) ,  \label{Model 7} \\
0\leq \alpha \leq \beta \leq 1,\;\;\frac{1}{2}\leq \beta \leq \frac{1+\alpha 
}{2},\;\;\frac{a_{3}}{a_{2}}\leq \frac{b_{2}}{b_{1}}\leq a_{2}\frac{1}{%
\left\vert \cos \left( \beta \pi \right) \right\vert },  \notag
\label{Model 7 - tdr} \\
\frac{a_{3}}{a_{2}}\leq \frac{a_{2}}{2\cos ^{2}\left( \beta \pi \right) }%
\left( 1-\sqrt{1-\frac{4a_{3}\cos ^{2}\left( \beta \pi \right) }{a_{2}^{2}}}%
\right) \leq \frac{b_{2}}{b_{1}}\leq \frac{a_{2}}{\left\vert \cos \left(
\beta \pi \right) \right\vert };  \notag
\end{gather}

\noindent \textbf{Model VIII}:%
\begin{gather}
\left( 1+\bar{a}_{1}\,{}_{0}\mathrm{D}_{t}^{\alpha }+\bar{a}_{2}\,{}_{0}%
\mathrm{D}_{t}^{2\alpha }\right) \sigma \left( t\right) =\left( b_{1}\,{}_{0}%
\mathrm{D}_{t}^{\alpha }+b_{2}\,{}_{0}\mathrm{D}_{t}^{2\alpha }\right)
\varepsilon \left( t\right) ,  \label{Model 8} \\
\frac{1}{2}\leq \alpha \leq 1,\;\;\frac{\bar{a}_{2}}{\bar{a}_{1}}\leq \frac{%
b_{2}}{b_{1}}\leq \bar{a}_{1}\frac{1}{\left\vert \cos \left( \alpha \pi
\right) \right\vert },  \notag  \label{Model 8 - tdr} \\
\frac{\bar{a}_{2}}{\bar{a}_{1}}\leq \frac{\bar{a}_{1}}{2\cos ^{2}\left(
\alpha \pi \right) }\left( 1-\sqrt{1-\frac{4\bar{a}_{2}\cos ^{2}\left(
\alpha \pi \right) }{\bar{a}_{1}^{2}}}\right) \leq \frac{b_{2}}{b_{1}}\leq 
\frac{\bar{a}_{1}}{\left\vert \cos \left( \alpha \pi \right) \right\vert }. 
\notag
\end{gather}

\section{Calculation of relaxation modulus and creep compliances \label%
{sr-cr-calc}}

The relaxation modulus (\ref{sigma-sr-eq}) and creep compliance (\ref%
{eps-cr-eq}) are calculated by the Laplace transform inversion formula 
\begin{equation}
f\left( t\right) =\mathcal{L}^{-1}\left[ \tilde{f}\left( s\right) \right]
\left( t\right) =\frac{1}{2\pi \mathrm{i}}\lim_{\substack{ R\rightarrow
\infty ,  \\ r\rightarrow 0}}\int_{\Gamma _{0}}\tilde{f}\left( s\right) 
\mathrm{e}^{st}\mathrm{d}s,  \label{LIF}
\end{equation}%
with $\Gamma _{0}$ being the Bromwich path, respectively applied to the
relaxation modulus in complex domain%
\begin{equation}
\tilde{\sigma}_{sr}\left( s\right) =\frac{1}{s}\frac{\Phi _{\varepsilon }(s)%
}{\Phi _{\sigma }(s)},\;\;\func{Re}s>p_{0}\geq 0,  \label{sigma-sr-tilda}
\end{equation}%
and creep compliance in complex domain%
\begin{equation}
\tilde{\varepsilon}_{cr}\left( s\right) =\frac{1}{s}\frac{\Phi _{\sigma }(s)%
}{\Phi _{\varepsilon }(s)},\;\;\func{Re}s>p_{0}\geq 0,  \label{eps-cr-tilda}
\end{equation}%
see also (\ref{sr-cr}), while the creep compliance (\ref{eps-cr}) is
obtained as%
\begin{eqnarray}
\varepsilon _{cr}\left( t\right) &=&\int_{0}^{t}f_{cr}\left( t^{\prime
}\right) \,\mathrm{d}t^{\prime },\;\text{with}  \notag \\
f_{cr}\left( t\right) &=&\mathcal{L}^{-1}\left[ \tilde{f}_{cr}\left(
s\right) \right] \left( t\right) =\frac{1}{\pi }\int_{0}^{\infty }\frac{%
K\left( \rho \right) }{\left\vert \Phi _{\varepsilon }\left( \rho \mathrm{e}%
^{\mathrm{i}\pi }\right) \right\vert ^{2}}\mathrm{e}^{-\rho t}\mathrm{d}\rho
\;\;\text{and}\;\;\tilde{f}_{cr}\left( s\right) =\frac{\Phi _{\sigma }(s)}{%
\Phi _{\varepsilon }(s)}  \label{fi-sigma/eps}
\end{eqnarray}%
from the creep compliance in complex domain (\ref{eps-cr-tilda}) by the use
of the Laplace transform inversion formula (\ref{LIF}).

\begin{figure}[tbph]
\centering
\includegraphics[scale=0.7]{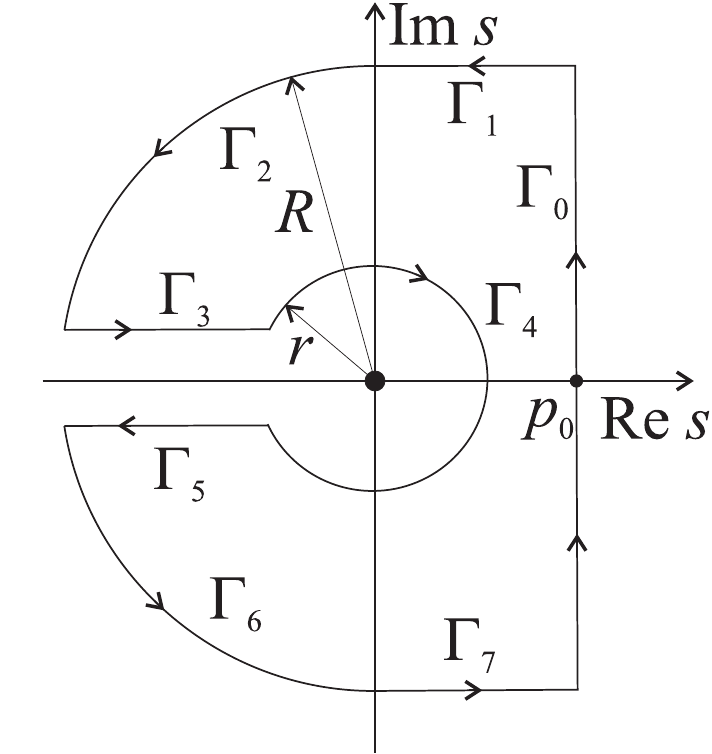}
\caption{Contour $\Gamma .$}
\label{fig-HK}
\end{figure}

Assuming $\left( A1\right) ,$ the Cauchy integral theorem%
\begin{equation}
\oint_{\Gamma }\tilde{f}\left( s\right) \mathrm{e}^{st}\mathrm{d}s=0,
\label{kif}
\end{equation}%
where $\Gamma $ is a closed curve containing the Bromwich path $\Gamma _{0},$
chosen as in Figure \ref{fig-HK}, is used in order to calculate the inverse
Laplace transform by (\ref{LIF}). 

The integrals along contours $\Gamma _{3}$ (parametrized by $s=\rho \mathrm{e%
}^{\mathrm{i}\pi },$ $\rho \in \left( r,R\right) $) and $\Gamma _{5}$
(parametrized by $s=\rho \mathrm{e}^{-\mathrm{i}\pi },$ $\rho \in \left(
r,R\right) $) in the case of relaxation modulus (\ref{sigma-sr-eq}) read 
\begin{eqnarray*}
\lim_{\substack{ R\rightarrow \infty ,  \\ r\rightarrow 0}}\int_{\Gamma _{3}}%
\tilde{\sigma}_{sr}\left( s\right) \mathrm{e}^{st}\mathrm{d}s
&=&\int_{\infty }^{0}\frac{1}{\rho \mathrm{e}^{\mathrm{i}\pi }}\frac{\Phi
_{\varepsilon }\left( \rho \mathrm{e}^{\mathrm{i}\pi }\right) }{\Phi
_{\sigma }\left( \rho \mathrm{e}^{\mathrm{i}\pi }\right) }\mathrm{e}^{-\rho
t}\mathrm{e}^{\mathrm{i}\pi }\mathrm{d}\rho =-\int_{0}^{\infty }\frac{\Phi
_{\varepsilon }\left( \rho \mathrm{e}^{\mathrm{i}\pi }\right) }{\Phi
_{\sigma }\left( \rho \mathrm{e}^{\mathrm{i}\pi }\right) }\frac{\mathrm{e}%
^{-\rho t}}{\rho }\mathrm{d}\rho , \\
\lim_{\substack{ R\rightarrow \infty ,  \\ r\rightarrow 0}}\int_{\Gamma _{5}}%
\tilde{\sigma}_{sr}\left( s\right) \mathrm{e}^{st}\mathrm{d}s
&=&\int_{0}^{\infty }\frac{1}{\rho \mathrm{e}^{-\mathrm{i}\pi }}\frac{\Phi
_{\varepsilon }\left( \rho \mathrm{e}^{-\mathrm{i}\pi }\right) }{\Phi
_{\sigma }\left( \rho \mathrm{e}^{-\mathrm{i}\pi }\right) }\mathrm{e}^{-\rho
t}\mathrm{e}^{-\mathrm{i}\pi }\mathrm{d}\rho =\int_{0}^{\infty }\frac{\bar{%
\Phi}_{\varepsilon }\left( \rho \mathrm{e}^{\mathrm{i}\pi }\right) }{\bar{%
\Phi}_{\sigma }\left( \rho \mathrm{e}^{\mathrm{i}\pi }\right) }\frac{\mathrm{%
e}^{-\rho t}}{\rho }\mathrm{d}\rho ,
\end{eqnarray*}%
while for the creep compliance in the form (\ref{eps-cr-eq}), the integrals
are%
\begin{eqnarray*}
\lim_{\substack{ R\rightarrow \infty ,  \\ r\rightarrow 0}}\int_{\Gamma _{3}}%
\tilde{\varepsilon}_{cr}\left( s\right) \mathrm{e}^{st}\mathrm{d}s
&=&\int_{\infty }^{0}\frac{1}{\rho \mathrm{e}^{\mathrm{i}\pi }}\frac{\Phi
_{\sigma }\left( \rho \mathrm{e}^{\mathrm{i}\pi }\right) }{\Phi
_{\varepsilon }\left( \rho \mathrm{e}^{\mathrm{i}\pi }\right) }\mathrm{e}%
^{-\rho t}\mathrm{e}^{\mathrm{i}\pi }\mathrm{d}\rho =-\int_{0}^{\infty }%
\frac{\Phi _{\sigma }\left( \rho \mathrm{e}^{\mathrm{i}\pi }\right) }{\Phi
_{\varepsilon }\left( \rho \mathrm{e}^{\mathrm{i}\pi }\right) }\frac{\mathrm{%
e}^{-\rho t}}{\rho }\mathrm{d}\rho , \\
\lim_{\substack{ R\rightarrow \infty ,  \\ r\rightarrow 0}}\int_{\Gamma _{5}}%
\tilde{\varepsilon}_{cr}\left( s\right) \mathrm{e}^{st}\mathrm{d}s
&=&\int_{0}^{\infty }\frac{1}{\rho \mathrm{e}^{-\mathrm{i}\pi }}\frac{\Phi
_{\sigma }\left( \rho \mathrm{e}^{-\mathrm{i}\pi }\right) }{\Phi
_{\varepsilon }\left( \rho \mathrm{e}^{-\mathrm{i}\pi }\right) }\mathrm{e}%
^{-\rho t}\mathrm{e}^{-\mathrm{i}\pi }\mathrm{d}\rho =\int_{0}^{\infty }%
\frac{\bar{\Phi}_{\sigma }\left( \rho \mathrm{e}^{\mathrm{i}\pi }\right) }{%
\bar{\Phi}_{\varepsilon }\left( \rho \mathrm{e}^{\mathrm{i}\pi }\right) }%
\frac{\mathrm{e}^{-\rho t}}{\rho }\mathrm{d}\rho ,
\end{eqnarray*}%
and for the creep compliance in the form (\ref{eps-cr}) their form is%
\begin{eqnarray*}
\lim_{\substack{ R\rightarrow \infty ,  \\ r\rightarrow 0}}\int_{\Gamma _{3}}%
\tilde{f}_{cr}\left( s\right) \mathrm{e}^{st}\mathrm{d}s &=&\int_{\infty
}^{0}\frac{\Phi _{\sigma }\left( \rho \mathrm{e}^{\mathrm{i}\pi }\right) }{%
\Phi _{\varepsilon }\left( \rho \mathrm{e}^{\mathrm{i}\pi }\right) }\mathrm{e%
}^{-\rho t}\mathrm{e}^{\mathrm{i}\pi }\mathrm{d}\rho =\int_{0}^{\infty }%
\frac{\Phi _{\sigma }\left( \rho \mathrm{e}^{\mathrm{i}\pi }\right) }{\Phi
_{\varepsilon }\left( \rho \mathrm{e}^{\mathrm{i}\pi }\right) }\mathrm{e}%
^{-\rho t}\mathrm{d}\rho , \\
\lim_{\substack{ R\rightarrow \infty ,  \\ r\rightarrow 0}}\int_{\Gamma _{5}}%
\tilde{f}_{cr}\left( s\right) \mathrm{e}^{st}\mathrm{d}s &=&\int_{0}^{\infty
}\frac{\Phi _{\sigma }\left( \rho \mathrm{e}^{-\mathrm{i}\pi }\right) }{\Phi
_{\varepsilon }\left( \rho \mathrm{e}^{-\mathrm{i}\pi }\right) }\mathrm{e}%
^{-\rho t}\mathrm{e}^{-\mathrm{i}\pi }\mathrm{d}\rho =-\int_{0}^{\infty }%
\frac{\bar{\Phi}_{\sigma }\left( \rho \mathrm{e}^{\mathrm{i}\pi }\right) }{%
\bar{\Phi}_{\varepsilon }\left( \rho \mathrm{e}^{\mathrm{i}\pi }\right) }%
\mathrm{e}^{-\rho t}\mathrm{d}\rho ,
\end{eqnarray*}%
where, according to (\ref{fiovi}), one has $\Phi _{\sigma }\left( \bar{s}%
\right) =\bar{\Phi}_{\sigma }\left( s\right) $ and $\Phi _{\varepsilon
}\left( \bar{s}\right) =\bar{\Phi}_{\varepsilon }\left( s\right) $ (bar
denotes the complex conjugation), so that%
\begin{eqnarray}
\lim_{\substack{ R\rightarrow \infty ,  \\ r\rightarrow 0}}\int_{\Gamma
_{3}\cup \Gamma _{5}}\tilde{\sigma}_{sr}\left( s\right) \mathrm{e}^{st}%
\mathrm{d}s &=&\int_{0}^{\infty }\left( \frac{\bar{\Phi}_{\varepsilon
}\left( \rho \mathrm{e}^{\mathrm{i}\pi }\right) }{\bar{\Phi}_{\sigma }\left(
\rho \mathrm{e}^{\mathrm{i}\pi }\right) }-\frac{\Phi _{\varepsilon }\left(
\rho \mathrm{e}^{\mathrm{i}\pi }\right) }{\Phi _{\sigma }\left( \rho \mathrm{%
e}^{\mathrm{i}\pi }\right) }\right) \frac{\mathrm{e}^{-\rho t}}{\rho }%
\mathrm{d}\rho =-2\mathrm{i}\int_{0}^{\infty }\frac{K\left( \rho \right) }{%
\left\vert \Phi _{\sigma }\left( \rho \mathrm{e}^{\mathrm{i}\pi }\right)
\right\vert ^{2}}\frac{\mathrm{e}^{-\rho t}}{\rho }\mathrm{d}\rho ,
\label{int-1.} \\
\lim_{\substack{ R\rightarrow \infty ,  \\ r\rightarrow 0}}\int_{\Gamma
_{3}\cup \Gamma _{5}}\tilde{\varepsilon}_{cr}\left( s\right) \mathrm{e}^{st}%
\mathrm{d}s &=&\int_{0}^{\infty }\left( \frac{\bar{\Phi}_{\sigma }\left(
\rho \mathrm{e}^{\mathrm{i}\pi }\right) }{\bar{\Phi}_{\varepsilon }\left(
\rho \mathrm{e}^{\mathrm{i}\pi }\right) }-\frac{\Phi _{\sigma }\left( \rho 
\mathrm{e}^{\mathrm{i}\pi }\right) }{\Phi _{\varepsilon }\left( \rho \mathrm{%
e}^{\mathrm{i}\pi }\right) }\right) \frac{\mathrm{e}^{-\rho t}}{\rho }%
\mathrm{d}\rho =2\mathrm{i}\int_{0}^{\infty }\frac{K\left( \rho \right) }{%
\left\vert \Phi _{\varepsilon }\left( \rho \mathrm{e}^{\mathrm{i}\pi
}\right) \right\vert ^{2}}\frac{\mathrm{e}^{-\rho t}}{\rho }\mathrm{d}\rho ,
\label{int-2.} \\
\lim_{\substack{ R\rightarrow \infty ,  \\ r\rightarrow 0}}\int_{\Gamma
_{3}\cup \Gamma _{5}}\tilde{f}_{cr}\left( s\right) \mathrm{e}^{st}\mathrm{d}%
s &=&\int_{0}^{\infty }\left( \frac{\Phi _{\sigma }\left( \rho \mathrm{e}^{%
\mathrm{i}\pi }\right) }{\Phi _{\varepsilon }\left( \rho \mathrm{e}^{\mathrm{%
i}\pi }\right) }-\frac{\bar{\Phi}_{\sigma }\left( \rho \mathrm{e}^{\mathrm{i}%
\pi }\right) }{\bar{\Phi}_{\varepsilon }\left( \rho \mathrm{e}^{\mathrm{i}%
\pi }\right) }\right) \mathrm{e}^{-\rho t}\mathrm{d}\rho =-2\mathrm{i}%
\int_{0}^{\infty }\frac{K\left( \rho \right) }{\left\vert \Phi _{\varepsilon
}\left( \rho \mathrm{e}^{\mathrm{i}\pi }\right) \right\vert ^{2}}\mathrm{e}%
^{-\rho t}\mathrm{d}\rho ,  \label{int-3.}
\end{eqnarray}%
respectively, with%
\begin{equation*}
K\left( \rho \right) =\frac{1}{2\mathrm{i}}\left( \bar{\Phi}_{\sigma }\left(
\rho \mathrm{e}^{\mathrm{i}\pi }\right) \Phi _{\varepsilon }\left( \rho 
\mathrm{e}^{\mathrm{i}\pi }\right) -\Phi _{\sigma }\left( \rho \mathrm{e}^{%
\mathrm{i}\pi }\right) \bar{\Phi}_{\varepsilon }\left( \rho \mathrm{e}^{%
\mathrm{i}\pi }\right) \right)
\end{equation*}%
giving (\ref{K}). According to the Cauchy integral theorem (\ref{kif}), the
integrals (\ref{int-1.}) and (\ref{int-2.}) yield the second terms in
relaxation modulus (\ref{sigma-sr-eq}) and creep compliance (\ref{eps-cr-eq}%
), while the integral (\ref{int-3.}) yields function $f_{cr}$ in (\ref%
{fi-sigma/eps}), since the integrals along $\Gamma _{1},$ $\Gamma _{2},$ $%
\Gamma _{6},$ $\Gamma _{7}$ tend to zero as $R\rightarrow \infty $ and $%
r\rightarrow 0,$ and the integral along $\Gamma _{4}$ is: nonzero in cases
of (\ref{sigma-sr-eq}) and (\ref{eps-cr-eq}), and zero in the case of (\ref%
{fi-sigma/eps}).

The contour $\Gamma _{1}$\ is parametrized by $s=p+iR,$\ $p\in \left(
0,p_{0}\right) ,$\ with $R\rightarrow \infty ,$\ so that the integrals%
\begin{eqnarray*}
\int_{\Gamma _{1}}\tilde{\sigma}_{sr}\left( s\right) \mathrm{e}^{st}\mathrm{d%
}s &=&\int_{p_{0}}^{0}\frac{1}{p+\mathrm{i}R}\frac{\Phi _{\varepsilon
}\left( p+\mathrm{i}R\right) }{\Phi _{\sigma }(p+\mathrm{i}R)}\mathrm{e}%
^{\left( p+\mathrm{i}R\right) t}\mathrm{d}p, \\
\int_{\Gamma _{1}}\tilde{\varepsilon}_{cr}\left( s\right) \mathrm{e}^{st}%
\mathrm{d}s &=&\int_{p_{0}}^{0}\frac{1}{p+\mathrm{i}R}\frac{\Phi _{\sigma
}\left( p+\mathrm{i}R\right) }{\Phi _{\varepsilon }(p+\mathrm{i}R)}\mathrm{e}%
^{\left( p+\mathrm{i}R\right) t}\mathrm{d}p, \\
\int_{\Gamma _{1}}\tilde{f}_{cr}\left( s\right) \mathrm{e}^{st}\mathrm{d}s
&=&\int_{p_{0}}^{0}\frac{\Phi _{\sigma }\left( p+\mathrm{i}R\right) }{\Phi
_{\varepsilon }(p+\mathrm{i}R)}\mathrm{e}^{\left( p+\mathrm{i}R\right) t}%
\mathrm{d}p
\end{eqnarray*}%
are estimated as%
\begin{eqnarray*}
\left\vert \int_{\Gamma _{1}}\tilde{\sigma}_{sr}\left( s\right) \mathrm{e}%
^{st}\mathrm{d}s\right\vert &\leq &\int_{0}^{p_{0}}\frac{1}{\left\vert p+%
\mathrm{i}R\right\vert }\left\vert \frac{\Phi _{\varepsilon }\left( p+%
\mathrm{i}R\right) }{\Phi _{\sigma }(p+\mathrm{i}R)}\right\vert \mathrm{e}%
^{pt}\mathrm{d}p, \\
\left\vert \int_{\Gamma _{1}}\tilde{\varepsilon}_{cr}\left( s\right) \mathrm{%
e}^{st}\mathrm{d}s\right\vert &\leq &\int_{0}^{p_{0}}\frac{1}{\left\vert p+%
\mathrm{i}R\right\vert }\left\vert \frac{\Phi _{\sigma }\left( p+\mathrm{i}%
R\right) }{\Phi _{\varepsilon }(p+\mathrm{i}R)}\right\vert \mathrm{e}^{pt}%
\mathrm{d}p, \\
\left\vert \int_{\Gamma _{1}}\tilde{f}_{cr}\left( s\right) \mathrm{e}^{st}%
\mathrm{d}s\right\vert &\leq &\int_{0}^{p_{0}}\left\vert \frac{\Phi _{\sigma
}\left( p+\mathrm{i}R\right) }{\Phi _{\varepsilon }(p+\mathrm{i}R)}%
\right\vert \mathrm{e}^{pt}\mathrm{d}p.
\end{eqnarray*}%
Assuming $s=\rho e^{\mathrm{i}\varphi },$\ since $R\rightarrow \infty ,$\
one obtains $\rho =\sqrt{p^{2}+R^{2}}\sim R$\ and $\varphi =\arctan \frac{R}{%
p}\sim \frac{\pi }{2},$\ the previous expressions become%
\begin{eqnarray*}
\lim_{R\rightarrow \infty }\left\vert \int_{\Gamma _{1}}\tilde{\sigma}%
_{sr}\left( s\right) \mathrm{e}^{st}\mathrm{d}s\right\vert &\leq
&\lim_{R\rightarrow \infty }\int_{0}^{p_{0}}\frac{1}{R}\left\vert \frac{\Phi
_{\varepsilon }\left( R\mathrm{e}^{\mathrm{i}\frac{\pi }{2}}\right) }{\Phi
_{\sigma }(R\mathrm{e}^{\mathrm{i}\frac{\pi }{2}})}\right\vert \mathrm{e}%
^{pt}\mathrm{d}p=0, \\
\lim_{R\rightarrow \infty }\left\vert \int_{\Gamma _{1}}\tilde{\varepsilon}%
_{cr}\left( s\right) \mathrm{e}^{st}\mathrm{d}s\right\vert &\leq
&\lim_{R\rightarrow \infty }\int_{0}^{p_{0}}\frac{1}{R}\left\vert \frac{\Phi
_{\sigma }\left( R\mathrm{e}^{\mathrm{i}\frac{\pi }{2}}\right) }{\Phi
_{\varepsilon }(R\mathrm{e}^{\mathrm{i}\frac{\pi }{2}})}\right\vert \mathrm{e%
}^{pt}\mathrm{d}p=0, \\
\lim_{R\rightarrow \infty }\left\vert \int_{\Gamma _{1}}\tilde{f}_{cr}\left(
s\right) \mathrm{e}^{st}\mathrm{d}s\right\vert &\leq &\lim_{R\rightarrow
\infty }\int_{0}^{p_{0}}\left\vert \frac{\Phi _{\sigma }\left( R\mathrm{e}^{%
\mathrm{i}\frac{\pi }{2}}\right) }{\Phi _{\varepsilon }(R\mathrm{e}^{\mathrm{%
i}\frac{\pi }{2}})}\right\vert \mathrm{e}^{pt}\mathrm{d}p=0,
\end{eqnarray*}%
due to assumptions $\left( A2\right) ,$ $\left( A3\right) ,$ and $\left(
A4\right) ,$ respectively. Analogously, it can be proved that the integral
along $\Gamma _{7}$ tends to zero as well.

The integrals along contour $\Gamma _{2},$ parametrized by $s=R\mathrm{e}^{%
\mathrm{i}\varphi },$ $\varphi \in \left( \frac{\pi }{2},\pi \right) ,$ with 
$R\rightarrow \infty ,$ are%
\begin{eqnarray*}
\int_{\Gamma _{2}}\tilde{\sigma}_{sr}\left( s\right) \mathrm{e}^{st}\mathrm{d%
}s &=&\int_{\frac{\pi }{2}}^{\pi }\frac{1}{R\mathrm{e}^{\mathrm{i}\varphi }}%
\frac{\Phi _{\varepsilon }(R\mathrm{e}^{\mathrm{i}\varphi })}{\Phi _{\sigma
}(R\mathrm{e}^{\mathrm{i}\varphi })}\mathrm{e}^{Rt\mathrm{e}^{\mathrm{i}%
\varphi }}\mathrm{i}R\mathrm{e}^{\mathrm{i}\varphi }\mathrm{d}\varphi , \\
\int_{\Gamma _{2}}\tilde{\varepsilon}_{cr}\left( s\right) \mathrm{e}^{st}%
\mathrm{d}s &=&\int_{\frac{\pi }{2}}^{\pi }\frac{1}{R\mathrm{e}^{\mathrm{i}%
\varphi }}\frac{\Phi _{\sigma }(R\mathrm{e}^{\mathrm{i}\varphi })}{\Phi
_{\varepsilon }(R\mathrm{e}^{\mathrm{i}\varphi })}\mathrm{e}^{Rt\mathrm{e}^{%
\mathrm{i}\varphi }}\mathrm{i}R\mathrm{e}^{\mathrm{i}\varphi }\mathrm{d}%
\varphi , \\
\int_{\Gamma _{2}}\tilde{f}_{cr}\left( s\right) \mathrm{e}^{st}\mathrm{d}s
&=&\int_{\frac{\pi }{2}}^{\pi }\frac{\Phi _{\sigma }(R\mathrm{e}^{\mathrm{i}%
\varphi })}{\Phi _{\varepsilon }(R\mathrm{e}^{\mathrm{i}\varphi })}\mathrm{e}%
^{Rt\mathrm{e}^{\mathrm{i}\varphi }}\mathrm{i}R\mathrm{e}^{\mathrm{i}\varphi
}\mathrm{d}\varphi
\end{eqnarray*}%
respectively, yielding the estimates%
\begin{eqnarray*}
\left\vert \int_{\Gamma _{2}}\tilde{\sigma}_{sr}\left( s\right) \mathrm{e}%
^{st}\mathrm{d}s\right\vert &\leq &\int_{\frac{\pi }{2}}^{\pi }\left\vert 
\frac{\Phi _{\varepsilon }(R\mathrm{e}^{\mathrm{i}\varphi })}{\Phi _{\sigma
}(R\mathrm{e}^{\mathrm{i}\varphi })}\right\vert \mathrm{e}^{Rt\cos \varphi }%
\mathrm{d}\varphi =0, \\
\left\vert \int_{\Gamma _{2}}\tilde{\varepsilon}_{cr}\left( s\right) \mathrm{%
e}^{st}\mathrm{d}s\right\vert &\leq &\int_{\frac{\pi }{2}}^{\pi }\left\vert 
\frac{\Phi _{\sigma }(R\mathrm{e}^{\mathrm{i}\varphi })}{\Phi _{\varepsilon
}(R\mathrm{e}^{\mathrm{i}\varphi })}\right\vert \mathrm{e}^{Rt\cos \varphi }%
\mathrm{d}\varphi =0, \\
\left\vert \int_{\Gamma _{2}}\tilde{f}_{cr}\left( s\right) \mathrm{e}^{st}%
\mathrm{d}s\right\vert &\leq &\int_{\frac{\pi }{2}}^{\pi }R\left\vert \frac{%
\Phi _{\sigma }(R\mathrm{e}^{\mathrm{i}\varphi })}{\Phi _{\varepsilon }(R%
\mathrm{e}^{\mathrm{i}\varphi })}\right\vert \mathrm{e}^{Rt\cos \varphi }%
\mathrm{d}\varphi =0.
\end{eqnarray*}%
due to assumptions $\left( A2\right) ,$ $\left( A3\right) ,$ and $\left(
A4\right) ,$ respectively. By the similar arguments, the integral along $%
\Gamma _{6}$ tends to zero as well.

Parametrization of the contour $\Gamma _{4}$ is $s=r\mathrm{e}^{\mathrm{i}%
\varphi },$ $\varphi \in \left( -\pi ,\pi \right) ,$ with $r\rightarrow 0,$
so that%
\begin{eqnarray*}
\int_{\Gamma _{4}}\tilde{\sigma}_{sr}\left( s\right) \mathrm{e}^{st}\mathrm{d%
}s &=&\int_{\pi }^{-\pi }\frac{1}{r\mathrm{e}^{\mathrm{i}\varphi }}\frac{%
\Phi _{\varepsilon }(r\mathrm{e}^{\mathrm{i}\varphi })}{\Phi _{\sigma }(r%
\mathrm{e}^{\mathrm{i}\varphi })}\mathrm{e}^{rt\mathrm{e}^{\mathrm{i}\varphi
}}\mathrm{i}r\mathrm{e}^{\mathrm{i}\varphi }\mathrm{d}\varphi , \\
\int_{\Gamma _{4}}\tilde{\varepsilon}_{cr}\left( s\right) \mathrm{e}^{st}%
\mathrm{d}s &=&\int_{\pi }^{-\pi }\frac{1}{r\mathrm{e}^{\mathrm{i}\varphi }}%
\frac{\Phi _{\sigma }(r\mathrm{e}^{\mathrm{i}\varphi })}{\Phi _{\varepsilon
}(r\mathrm{e}^{\mathrm{i}\varphi })}\mathrm{e}^{rt\mathrm{e}^{\mathrm{i}%
\varphi }}\mathrm{i}r\mathrm{e}^{\mathrm{i}\varphi }\mathrm{d}\varphi , \\
\int_{\Gamma _{4}}\tilde{f}_{cr}\left( s\right) \mathrm{e}^{st}\mathrm{d}s
&=&\int_{\pi }^{-\pi }\frac{\Phi _{\sigma }(r\mathrm{e}^{\mathrm{i}\varphi })%
}{\Phi _{\varepsilon }(r\mathrm{e}^{\mathrm{i}\varphi })}\mathrm{e}^{rt%
\mathrm{e}^{\mathrm{i}\varphi }}\mathrm{i}r\mathrm{e}^{\mathrm{i}\varphi }%
\mathrm{d}\varphi
\end{eqnarray*}%
respectively, yield%
\begin{eqnarray*}
\lim_{r\rightarrow 0}\int_{\Gamma _{4}}\tilde{\sigma}_{sr}\left( s\right) 
\mathrm{e}^{st}\mathrm{d}s &=&-\mathrm{i\,}\lim_{r\rightarrow 0}\int_{-\pi
}^{\pi }\frac{\Phi _{\varepsilon }(r\mathrm{e}^{\mathrm{i}\varphi })}{\Phi
_{\sigma }(r\mathrm{e}^{\mathrm{i}\varphi })}\mathrm{d}\varphi =-2\pi 
\mathrm{i}\sigma _{sr}^{\left( e\right) }, \\
\lim_{r\rightarrow 0}\int_{\Gamma _{4}}\tilde{\varepsilon}_{cr}\left(
s\right) \mathrm{e}^{st}\mathrm{d}s &=&-\mathrm{i\,}\lim_{r\rightarrow
0}\int_{-\pi }^{\pi }\frac{\Phi _{\sigma }(r\mathrm{e}^{\mathrm{i}\varphi })%
}{\Phi _{\varepsilon }(r\mathrm{e}^{\mathrm{i}\varphi })}\mathrm{d}\varphi
=-2\pi \mathrm{i}\varepsilon _{cr}^{\left( e\right) }, \\
\lim_{r\rightarrow 0}\left\vert \int_{\Gamma _{4}}\tilde{f}_{cr}\left(
s\right) \mathrm{e}^{st}\mathrm{d}s\right\vert &\leq &\lim_{r\rightarrow
0}\int_{-\pi }^{\pi }r\left\vert \frac{\Phi _{\sigma }(r\mathrm{e}^{\mathrm{i%
}\varphi })}{\Phi _{\varepsilon }(r\mathrm{e}^{\mathrm{i}\varphi })}%
\right\vert \mathrm{d}\varphi =0,
\end{eqnarray*}%
due to (\ref{sigma-sr-e}), (\ref{eps-cr-e}), and assumption $\left(
A4\right) .$

\section*{Acknowledgment}

This work is supported by the Serbian Ministry of Education, Science and
Technological Development under grants $174005$ and $174024,$ by the
Provincial Secretariat for Higher Education and Scientific Research under
grant $142-451-2102/2019$, as well as by FWO Odysseus project of Michael
Ruzhansky.


\end{document}